\documentclass[twocolumn]{autart}    

\usepackage{graphicx}          

\usepackage{amsmath}

\usepackage{amsfonts}
\usepackage{mathabx}
\usepackage{color}
\usepackage{epstopdf} 
\usepackage{tikz}
\usepackage{mathtools}

\newcommand*{\red}{\textcolor{black}}

\newtheorem{proposition}{Proposition}

\newtheorem{remark}{Remark}
\newtheorem{lemma}{Lemma}

\newtheorem{theorem}{Theorem}

\newcommand{\bg}{\mathbf g}
\newcommand{\bm}{\mathbf m}

\newcommand{\mN}{\mathcal{N}}

\newcommand{\beq}{\begin{equation}}
\newcommand{\eeq}{\end{equation}}
\newcommand{\beqs}{\begin{equation*}}
\newcommand{\eeqs}{\end{equation*}}
\newcommand{\beqr}{\begin{eqnarray}}
\newcommand{\eeqr}{\end{eqnarray}}
\newcommand{\beqrs}{\begin{eqnarray*}}
	\newcommand{\eeqrs}{\end{eqnarray*}}
\newcommand*{\blue}{\textcolor{black}}

\DeclareMathOperator*{\argmin}{arg\,min}
\DeclareMathOperator*{\argmax}{arg\,max}
\usepackage{algorithm}
\newcommand{\norm}[1]{\| #1 \|}

\usepackage{cite}

\begin{document}

\begin{frontmatter}

\title{Learning linear modules in a dynamic network using regularized kernel-based methods\thanksref{footnoteinfo}} 

\thanks[footnoteinfo]{\blue{Paper submitted to Automatica, 12 May 2020; revised version 29 December 2020. Final version 6 January 2021.} This project has received funding from the European Research Council (ERC), Advanced Research Grant SYSDYNET, under the European Union’s Horizon 2020 research and innovation programme (Grant Agreement No. 694504).}

\author[Paestum]{Karthik R. Ramaswamy}\ead{k.r.ramaswamy@tue.nl},    
\author[Paestum]{Giulio Bottegal}\ead{giulio.bottegal@gmail.com},               
\author[Paestum]{Paul M.J. Van den Hof}\ead{p.m.j.vandenhof@tue.nl}  

\address[Paestum]{Department of Electrical Engineering, Eindhoven University of Technology, Eindhoven, The Netherlands}


\begin{abstract}
In order to identify one system (module) in an interconnected dynamic network, one typically has to solve a Multi-Input-Single-Output (MISO) identification problem that requires identification of all modules in the MISO setup. For application of a parametric identification method this would require estimating a large number of parameters, as well as an appropriate model order selection step for a possibly large scale MISO problem, thereby increasing the computational complexity of the identification algorithm to levels that are beyond feasibility. An alternative identification approach is presented employing regularized kernel-based methods. Keeping a parametric model for the module of interest, we model the impulse response of the remaining modules in the MISO structure as zero mean Gaussian processes (GP) with a covariance matrix (kernel) given by the first-order stable spline kernel, accounting for the noise model affecting the output of the target module and also for possible instability of systems in the MISO setup. Using an Empirical Bayes (EB) approach the target module parameters are estimated through an Expectation-Maximization (EM) algorithm with a substantially reduced computational complexity, while avoiding extensive model structure selection. Numerical simulations illustrate the potentials of the introduced method in comparison with the state-of-the-art techniques for local module identification. 
\end{abstract}

\end{frontmatter}

\section{Introduction}
Interconnected systems are becoming increasingly ubiquitous and 
\red{data-driven modeling problems in }
large-scale interconnected systems, known as dynamic networks, 
\red{is expected to become of paramount importance}
\blue{in different fields like robotics, smart grids, transportation systems, oil and gas reservoirs \cite{Mansoori&etal_IFAC:2014}, autonomous vehicle platooning \cite{Pimentel&etal_IFAC:2020}}. These networks can be considered as a set of measurable signals (the node signals) interconnected through linear dynamic systems and can be possibly driven by external excitation signals and/or process noise. \blue{Data-driven \red{modeling} methods for dynamic networks can be typically divided into three categories, namely finding the interconnection structure (topology) of the dynamic network \cite{Materassi10,Chiuso&Pillonetto_Autom:12,Shi&etal_ECC:19}, methods for full network identification, and methods for local module identification. Full network identification deals with the identification of the full network dynamics \cite{Haber&Verhaegen:TAC:14,Torres14,Weerts&etal_Autom:18_reducedrank,Weerts&etal_CDC:16,Zorzi&Chiuso:17}, including aspects of identifiability \cite{Goncalves&Warnick:08, Weerts&etal_Autom:18_identifiability, Hendrickx&Gevers&Bazanella_TAC:19,Bazanella&etal_CDC:17,vanWaarde&Tesi&Camlibel_NECSYS:18,Cheng&etal_CDC:19}, while local module identification deals with the identification of a specific module (system) of the network considering that the topology of the network is known \cite{VandenHof&etal_Autom:13,Dankers&etal_Autom:15,materassi2015identification,Dankers&etal_TAC:16,Linder&Enqvist_ijc:17,Ramaswamy&etal_CDC:18, Everitt&Bottegal&Hjalmarsson_Autom:18,Gevers&etal:sysid18,VandenHof&etal_CDC:19,Ramaswamy&etal_CDC:19,Ramaswamyetal_TAC:19,Materassi&Salapaka:19}.}

In this paper we focus on the local module identification problem. In \cite{VandenHof&etal_Autom:13,Dankers&etal_TAC:16}, the classical \emph{direct method} for closed loop identification \cite{Ljung:99} has been generalized to the framework of a dynamic network. Similarly, in \cite{Gevers&etal:sysid18,VandenHof&etal_Autom:13,Dankers&etal_TAC:16}, the indirect identification methods have been generalized to the dynamic network framework. A direct method to handle correlated process noise has been provided in \cite{VandenHof&etal_CDC:19, Ramaswamyetal_TAC:19}. A method that combines the frameworks of the direct and the indirect method by using additional excitation signals as predictor inputs has been introduced in \cite{Ramaswamy&etal_CDC:19}. Considering the effect of sensor noise in the measurements, the aforementioned setting has been generalized in \cite{Dankers&etal_Autom:15}. A \emph{simultaneous minimization of the prediction error} approach is introduced in \cite{Gunes&etal_IFAC:14} for identifying the target module in a dynamic network with only sensor noise. This method has been extended to a Bayesian setting in \cite{Everitt&Bottegal&Hjalmarsson_Autom:18}, where regularized kernel-based methods are used to decrease the variance of the estimated target module.

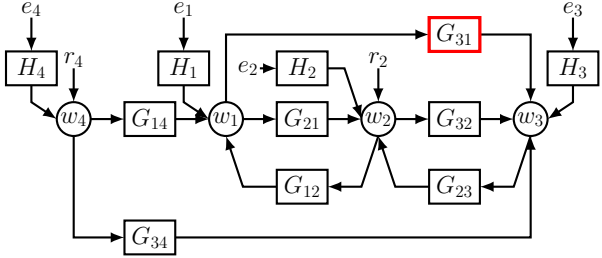
\begin{figure}
	\centering
	\usetikzlibrary{arrows}
	\begin{tikzpicture}[thick,scale=0.45, every node/.style={scale=0.45}]
	
	\draw  (-4.5,4.5) rectangle (-3,3.5) node[pos = 0.5] {\LARGE $G_{21}$};
	\draw [-latex](-5.5,4) -- (-4.5,4);
	\draw [-latex] (-6,4) ellipse (0.5 and 0.5) node{\LARGE $w_1$};
	\draw [-latex](-7.5,4) -- (-6.5,4);
	\draw [-latex] (-9,4.5) rectangle (-7.5,3.5) node[pos = 0.5] {\LARGE $G_{14}$};
	\draw [-latex](-10,4) -- (-9,4);
	\draw [-latex] (-10.5,4) ellipse (0.5 and 0.5) node{\LARGE $w_4$};
	\draw [-latex](-3,4) -- (-2,4);
	\draw [-latex] (-1.5,4) ellipse (0.5 and 0.5) node{\LARGE $w_2$};
	\draw [-latex](-1,4) -- (0,4);
	\draw [-latex] (0,4.5) rectangle (1.5,3.5) node[pos = 0.5] {\LARGE $G_{32}$};
	\draw [-latex](1.5,4) -- (2.5,4);
	\draw [-latex] (3,4) ellipse (0.5 and 0.5) node{\LARGE $w_3$};
	\draw [-latex](-10.5,5.5) node at (-10.5,5.75) {\LARGE $r_4$}-- (-10.5,4.5);
	\draw [-latex](-1.5,5.5) node at (-1.5,5.75) {\LARGE $r_2$}-- (-1.5,4.5);
	\draw [-latex] (-4.5,2.5) rectangle (-3,1.5) node[pos = 0.5] {\LARGE $G_{12}$};
	\draw [-latex] (0,2.5) rectangle (1.5,1.5) node[pos = 0.5] {\LARGE $G_{23}$};
	\draw [-latex](-1.5,3.5) -- (-2,2) -- (-3,2);
	\draw [-latex](3,3.5) -- (2.5,2) -- (1.5,2);
	\draw [-latex](0,2) -- (-1,2) -- (-1.5,3.5);
	\draw [-latex](-4.5,2) -- (-5.5,2) -- (-6,3.5);
	\draw [-latex] (-9,1) rectangle (-7.5,0) node[pos = 0.5] {\LARGE $G_{34}$};
	\draw [-latex](-10.5,3.5) -- (-10.5,0.5) -- (-9,0.5);
	\draw [-latex](-7.5,0.5) -- (3,0.5) -- (3,3.5);
	\draw [-latex] (-12.5,6) rectangle (-11,5) node[pos = 0.5] {\LARGE $H_{4}$};
	\draw [-latex] (0,7) rectangle (1.5,6) node[pos = 0.5] {\LARGE $G_{31}$};
	\draw [-latex](-6,4.5) -- (-6,6.5) -- (0,6.5);
	\draw [-latex](1.5,6.5) -- (3,6.5) -- (3,4.5);
	\draw [-latex] (-4.5,6) rectangle (-3,5) node[pos = 0.5] {\LARGE $H_{2}$};
	\draw [-latex] (-6.5,6) rectangle (-8,5) node[pos = 0.5] {\LARGE $H_{1}$};
	\draw [-latex] (3.5,6) rectangle (5,5) node[pos = 0.5] {\LARGE $H_{3}$};
	\draw [-latex](-11.75,5) -- (-11.75,4.5) -- (-11,4);
	\draw [-latex](-7.25,5) -- (-7.25,4.5) -- (-6.5,4);
	\draw [-latex](-3,5.5) -- (-2.5,5.5) -- (-2,4);
	\draw [-latex](-5,5.5) node at (-5.35,5.5) {\LARGE $e_2$} -- (-4.5,5.5);
	\draw [-latex](-11.75,7) node at (-11.75,7.25) {\LARGE $e_4$}-- (-11.75,6);
	\draw [-latex](-7.25,7) node at (-7.25,7.25) {\LARGE $e_1$} -- (-7.25,6);
	\draw [-latex](4.25,5) -- (4.25,4.5) -- (3.5,4);
	\draw [-latex](4.25,7) node at (4.25,7.25) {\LARGE $e_3$} -- (4.25,6);
	\draw [-latex, red, very thick] (0,7) rectangle (1.5,6);
	\end{tikzpicture}
	\caption{Network example with 4 internal nodes, 2 reference signals and a noise sources at each node.}
	\label{fig:dynnet_Ex_wnoise1}
\end{figure}

In this paper we aim at improving the performance of the direct method for dynamic networks, since the direct method exploits both the external excitation signals and noise signals for data informativity. Assuming a known topology of the network, in \cite{VandenHof&etal_Autom:13} it was shown that, in order to identify a given module of interest using the direct method, we have to formulate a multi-input single-output (MISO) identification problem where the inputs of the MISO setup correspond to the inputs of all modules of the network sharing the same output with the module of interest (see Sec. \ref{sec:Direct_Id} for details). A relaxed setup has been provided in \cite{Dankers&etal_TAC:16}, where the MISO setup contains only a subset of the above mentioned inputs. This implies that, in both the approaches, to avoid possible bias in the parameter estimates, one has to identify all the modules constituting the MISO structure, bringing in the problem a possibly high number of parameters to be estimated that are of no primal interest to the experimenter. For example, considering the network in Figure \ref{fig:dynnet_Ex_wnoise1} with the target module of interest for identification being $G_{31}$, one has to identify $G_{31}$, $G_{32}$ and $G_{34}$. Adding to this, a model order selection step needs to be performed to select the number of parameters for each module using complexity criteria like AIC, BIC, or cross validation \cite{Ljung:99}. For this, it is required to test a number of combination of candidate model orders that increases exponentially with the number of models in the MISO structure, making the model order selection step computationally infeasible (e.g., for 5 modules with FIR model structure and orders from 1 to 5, one has to test $5^5$ possible combinations). More importantly, if any of the modules constituting the MISO structure is unstable, the prediction error identification approaches available from the literature cannot be used, since the predictors are unstable. We stress the presence of unstable modules is compatible with stable input-output dynamics in a network. For example, in the network of Figure \ref{fig:dynnet_Ex_wnoise1} the effect of unstable modules in $G_{31}$ and/or $G_{32}$ could be canceled by suitable controllers $G_{23}$ and/or $G_{12}$.

In this paper, we address the aforementioned problems developing an identification method based on non-parametric regularized kernel-based methods that
\begin{itemize}
    \item identifies a local module through a direct approach, exploiting both the external excitation signals and the disturbance signals for data informativity,
	\item avoids the complexity of model order selection for large-scale problems,
	\item reduces the number of nuisance parameters that need to be estimated in local module identification, and
	\item can be used irrespective of the stability of the modules in the MISO structure, with no need of prior information on possible unstable modules.
\end{itemize}
In \cite{Ramaswamy&etal_CDC:18}, a method to improve the performance of the direct method for dynamic networks based on non-parametric regularized kernel based methods has been introduced. Even though the method in \cite{Ramaswamy&etal_CDC:18} achieves the first three above mentioned objectives, it does not achieve the fourth and cannot be used under the presence of unstable modules in the MISO structure.
In the current paper, building upon the preliminary work of \cite{Ramaswamy&etal_CDC:18}, we provide a different and unified framework to identify the module of interest, which does not depend on the stability of the modules in the MISO structure.

In order to develop this method, we build on the following approach. We keep a parametric model for the target module of interest in order to have an accurate description of its dynamics. The impulse responses of the remaining modules in the MISO structure are modeled as zero mean Gaussian Processes (GP), with covariance (or kernel) given by the first-order stable spline kernel \cite{Chenetal_Autom:12}, \cite{Pillonettoetal_Autom:14}, which encodes stability and smoothness of the processes. However, we need to handle the prior inclusion of stability property using kernel-based methods under the presence of unstable modules and also incorporate process noise modeling in our framework to avoid increased bias in the estimated target module. We do this by appropriately rewriting the network dynamics.


Using the aforementioned approach, we obtain a Gaussian probabilistic description that depends on a vector of parameters $\eta$ containing the parameters of the module of interest, the variance of the output noise, and the hyperparamaters characterizing the stable spline kernel. Therefore, estimating $\eta$ provides the parameters of the target module. This is accomplished by using an Empirical Bayes (EB) approach \cite{Maritz&Lwin:1989}, where $\eta$ is estimated by maximizing the marginal likelihood of the data, which requires solving a nonlinear non-convex optimization problem. To this end, we use the Expectation-Maximization (EM) method \cite{Dempsteretal_JRSS:1977}, which provides a solution by iterating over simple sub-problems which  either admit analytical solutions or require solving scalar optimization problems. Numerical experiments performed on simulated dynamic networks show the potentials of the developed method in comparison with available classical methods.

This paper is organized as follows. In Section \ref{sec:prob stat}, the setup of the dynamic network is defined. Section \ref{sec:Direct_Id} provides a summary about the direct method and the extension of this framework using regularized kernel-based methods to end up in a marginal likelihood estimation problem is provided in Section \ref{sec:technique}. Next, we provide the approach and solution to the marginal likelihood problem using EM method. Section \ref{sec:Num_ex} provides the results of numerical simulations performed on simple dynamic networks, which is followed by the Conclusions. The technical proofs of all results are collected in the Appendix.

\section{Problem statement}\label{sec:prob stat}
Following the setup of \cite{VandenHof&etal_Autom:13}, we consider a dynamic network that is built up of $L$ scalar measurable internal variables or nodes $w_j(t)$, $j$ = $1, \dots, L$. The dynamic network is defined by the equation (time and frequency dependence is omitted below),
\begin{equation}\label{eq:networkmodel}
\begin{split}
\begin{bmatrix}
w_1 \\ w_2\\ \vdots \\ w_L
\end{bmatrix} &\!\!=\!\!
\begin{bmatrix}
\!\!0 & \!\!G_{12}^0(q)& \!\!\dots & \!\!G_{1L}^0(q)\\
\!\!G_{21}^0(q) & \!\!0& \!\!\dots & \!\!G_{2L}^0(q)\\
\!\!\vdots & \!\!\ddots & \!\!\ddots &\!\!\vdots \\
\!\!G_{L1}^0(q) & \!\!G_{L2}^0(q)& \!\!\dots & \!\!0\\
\end{bmatrix}\!\!\!\begin{bmatrix}
w_1 \\ w_2\\ \vdots \\ w_L
\end{bmatrix} \!\!+\!\!
\begin{bmatrix}
r_1 \\ r_2\\ \vdots \\ r_L
\end{bmatrix} \!\!+\!\!
\begin{bmatrix}
v_1 \\ v_2\\ \vdots \\ v_L
\end{bmatrix}\\
&= G^0(q)w(t) + r(t) + v(t)\\
\end{split}
\end{equation}
The representation in \eqref{eq:networkmodel} is an extension of the Dynamic Structure Function (DSF) representation \cite{Goncalves&Warnick:08}.
In the above equation,
\begin{itemize}
	\item $q^{-1}$ is the shift (delay) operator i.e. $q^{-1}u(t) = u(t - 1)$;
	\item $G_{jk}^0(q)$  is a strictly proper rational transfer function for $j = 1, \dots, L$ and $k = 1, \dots, L$;
	\item $v_j(t)$ is an unmeasured process noise entering node $w_j(t)$. It is a realization of a stationary stochastic process represented by $v_j(t) = H_j^0(q)e_j(t)$,  with $e_j(t)$ a Gaussian white noise process with unknown variance $\sigma_j^2$ and $H_j^0(q)$ a monic, stable and minimum phase filter;
	\item $r_j(t)$ is a measured external excitation signal entering node $w_j(t)$. In some nodes, it may be absent.
\end{itemize}
We assume that the dynamic network is stable, i.e. $(I - G^0(q))^{-1}$ is stable, and well posed (see \cite{VandenHof&etal_Autom:13} for details). Also we consider that the process noise $v_j(t)$ entering the node $w_j(t)$ is uncorrelated with the process noise entering any other node of the network.
We assume that we have collected $N$ measurements of the internal variables $\{w_k(t) \}_{t=1}^N$, $k = 1,\,\ldots,\,L$, and that we are interested in building a model of the module directly linking node $i$ to node $j$, that is $G_{ji}^0(q)$, using the measurements of the internal variables, and possibly $r$. To this end, we choose a parameterization of $G_{ji}^0(q)$, denoted as $G_{ji}(q,\theta)$, that describes the dynamics of the module of interest for a certain parameter vector $\theta_0 \in \mathbb{R}^{n_\theta}$.

We define $G_{jk}^0, k \in \mN_j$ and $H_j^0$ as rational transfer function such that $G_{jk}^0(q) = \frac{B_{jk}^0(q)}{F_{jk}^0(q)}$ and $H_{j}^0(q) = \frac{C_{j}^0(q)}{D_{j}^0(q)}$ where
\beqr
B_{jk}^0(q) &=& {b_{jk}^0}_1q^{-1} + \dots + {b_{jk}^0}_{n_b}q^{-n_{b_{jk}}}, \nonumber \\ \label{BAdefined1}
F_{jk}^0(q) &=& 1 + {f_{jk}^0}_1q^{-1} + \dots + {f_{jk}^0}_{n_f}q^{-n_{f_{jk}}},\\
C_{j}^0(q) &=& 1 + {c_{j}^0}_1q^{-1} + \dots + {c_{j}^0}_{n_c}q^{-n_{c_{j}}},  \nonumber \\
D_{j}^0(q) &=& 1 + {d_{j}^0}_1q^{-1} + \dots + {d_{j}^0}_{n_d}q^{-n_{d_{j}}}, \nonumber
\eeqr
are polynomials, and $n_{b_{jk}}, n_{f_{jk}}, n_{c_{jk}}, n_{d_{jk}}$ are positive integers, and $\mathcal{N}_j$ is the set of node indices $k$ such that $G_{jk} \nequiv 0$. We now expand the parameterization of $G_{ji}^0(q)$ as $G_{ji}(q,\theta) = \frac{B_{ji}(q,\theta_B)}{F_{ji}(q,\theta_F)} = \frac{B_{ji}(q,\theta_B)}{1 + \bar F_{ji}(q,\theta_F)}$ with $\theta = \begin{bmatrix}
\theta_B^\top & \theta_F^\top
\end{bmatrix}^\top$,
where $\theta_B$ and $\theta_F$ are the parameterized coefficients of polynomials $B_{ji}^0(q)$ and $F_{ji}^0(q)$ respectively as in Eq. \eqref{BAdefined1} (i.e. $\theta_B = [\begin{matrix} {b_{ji}}_1 & \dots & {b_{ji}}_{n_b}  \end{matrix}]^\top$ and $\theta_F = [\begin{matrix} {f_{ji}}_1 & \dots & {f_{ji}}_{n_f}  \end{matrix}]^\top$).

\section{The standard direct method}\label{sec:Direct_Id}
Following the definition of a dynamic network in the previous section, each scalar internal variable can be described as:
\begin{equation}\label{eq:singleblock}
 w_j(t) = \sum_{k \in \mathcal{N}_j} G_{jk}^0(q)w_k(t) + r_j(t) + v_j(t)
\end{equation}
The above equation represents a MISO structure and is the starting point of the methodology presented in this paper, which is based on extending the direct method \cite{VandenHof&etal_Autom:13}. In the standard direct method for dynamic networks \cite{VandenHof&etal_Autom:13}, we consider the one-step-ahead predictor \cite{Ljung:99} of $w_j(t)$:
\begin{equation*}
\resizebox{1.0\hsize}{!}{$
	\begin{split}
	\hat w_j(t|t-1;\theta) = &\big(1-({H_j})^{-1}(q,\theta)\big)w_j(t) + ({H_j})^{-1}(q,\theta) G_{ji}(q,\theta)w_i(t)\\
	&+ ({H_j})^{-1}(q,\theta)\big(\sum_{k \in \mathcal{N}_j\backslash\{i\}}G_{jk}(q,\theta)w_k(t) + r_j(t) \big)
	\end{split}$
}
\end{equation*}
which is a function of the parameter vector $\theta$. Not only the target module, but also the modules $G_{jk}^0(q)$, $k \in \mathcal{N}_j\backslash \{i\}$, and the noise model $H_j^0(q)$, are suitably parameterized with additional parameters. The parameter vector of interest $\theta$ is identified by minimizing the sum of the squared prediction error $\varepsilon_j(t) = w_j(t) - \hat w_j(t|t-1;\theta)$. We note that in this formulation, the prediction error depends also on the additional parameters entering the remaining modules and the noise model, which need to be identified to guarantee consistent estimates of $\theta$. Therefore, the total number of parameters may grow large if the cardinality of $\mathcal{N}_j$ is large, with a detrimental effect on the variance of the estimate of $\theta$ in the case where $N$ is not very large.

\section{The developed Empirical Bayes identification technique}\label{sec:technique}
We now discuss how to use regularized kernel-based methods to avoid parameterization of the additional modules (all modules except the target module) in the MISO structure. We define the following quantities:
$$
{S_j^0(q)} := 1 - (H_j^{0})^{-1}(q) \quad,\quad S_{jk}^0(q) := (H_j^{0})^{-1}G_{jk}^0(q) \,.
$$
Considering the above definitions, Eq. \eqref{eq:singleblock} can be re-written as
\begin{align}\label{eq:predictor}
w_j(t) &= \hat w_j(t|t-1) + e_j(t), \nonumber\\
&= {S_j^0(q)}w_j(t) + (1-S_j^0(q))(G_{ji}^0(q)w_i(t) + r_j(t)) \nonumber\\
&\quad+ \sum_{k \in \mathcal{N}_j\backslash\{i\}}S_{jk}^0(q)w_k(t) + e_j(t),
\end{align}
where we isolate the target module $G_{ji}^0(q)$. A main challenge when using kernel methods for LTI system identification is that typically a prior knowledge on the stability of the predictor filters in \eqref{eq:predictor} is imposed to reduce the MSE of the estimated impulse response of the system (see \cite{Pillonettoetal_Autom:14,Ramaswamy&etal_CDC:18}. When all systems (i.e. $G_{jk}, k \in \mN_j$) are stable, as assumed in \cite{Ramaswamy&etal_CDC:18}, the predictor filters in \eqref{eq:predictor} are stable and the setup in \eqref{eq:predictor} lends itself for kernel-based estimation of the predictor filters.
However, when some or all systems in the MISO structure are not stable, the imposition of prior knowledge on stability is not possible unless we suitably rewrite the network dynamics in \eqref{eq:singleblock}.     


\begin{proposition}\label{propnew1}
	Consider the network equation of the output node signal $w_{j}(t)$ in \eqref{eq:singleblock}. The network equation can be represented in an alternative way as\footnote{from now on superscript $^0$ is dropped for convenience.},
	\begin{align} \label{eq:43}
	&w_j(t) = {M_j(q)}w_j(t) - (1-M_j(q))\bar F_{ji}(q)w_j(t) \nonumber\\
	&+ (1\!-\!M_j(q)){B_{ji}(q)}w_i(t) +\!\!\!\!\!\! \sum_{k \in \mathcal{N}_j\backslash\{i\}} \!\!\!\!M_{jk}(q)w_k(t) \!+\! \bar e_j(t),
	\end{align}
	where $M_{\star}(q)$ are strictly proper predictor filters, $B_{ji}(q)$ and $\bar F_{ji}(q) = {-(1 - F_{ji}(q))}$ are stable polynomials representing $G_{ji}(q)$, and $\bar e_j(t)$ is a Gaussian white noise with variance $\bar \sigma_j^2$.
	
	{\bf Proof:} Collected in the appendix. The expressions for $M_{\star}(q)$ are provided in the appendix. \hfill\qed
\end{proposition}
Since all the predictor filters in the rewritten network dynamics are stable, this formulation lends itself to the Bayesian approach \cite{Ramaswamy&etal_CDC:18}, as described in the subsequent sections.

\subsection{Vector description of the dynamics}
In order to apply a kernel-based method to \eqref{eq:43}, we are going to formulate a vector description of the network dynamics for the available $N$ measurements. For notation purposes, we consider $N$-dimensional vectors $b_{ji}$ and $f_{ji}$ (which will also depend on $\theta$, although we will keep this dependence tacit) which are the parameterized coefficients of $B_{ji}(q, \theta_B)$ and $\bar F_{ji}(q, \theta_F)$ respectively stacked with zeros (i.e. $b_{ji} = [\begin{matrix} \theta_B^\top & \mathbf{0}^\top \end{matrix}]^\top$ and $f_{ji} = [\begin{matrix} \theta_F^\top & \mathbf{0}^\top \end{matrix}]^\top$). Similarly, we define the vector ${m_{k}}$, $k \in \mathcal N_j \backslash \{i\}$, and $m_j$ as the vectors containing the first $l$ coefficients of the impulse responses of ${M_{jk}}(q)$, $k \in \mathcal N_j \backslash \{i\}$, and $M_j(q)$, respectively. The integer $l$ is chosen large enough to ensure $m_{k}(l+1), m_{j}(l+1) \simeq 0$.

\begin{lemma}\label{propnew2}
	Let the vector notation for the node $w_j(t)$ be $w_j := \begin{bmatrix}
w_j(1) & \ldots & w_j(N)
\end{bmatrix}^T$. Considering the parameterization of $G_{ji}^0$, the network dynamics in \eqref{eq:43} can be represented in the vector form as:
	\begin{equation}\label{eq:44}
	w_j = \tilde W m_j + \mathbf W_{ji}g_{ji} + \sum_{k \in \mathcal{N}_j \backslash \{i \}}  W_{k} m_k + \bar e_j,
	\end{equation} where $\bg_{ji} = [\begin{matrix} b_{ji}^\top & f_{ji}^\top \end{matrix}]^\top$ and $\bar e_j$ is the vectorized noise. $\tilde W$, $\mathbf W_{ji}$ and $W_k$ are Toeplitz matrices constructed from measurements of the nodes in the MISO structure.
	
	{\bf Proof:} We denote by $W_k \in \mathbb{R}^{N\times l}$ the Toeplitz matrix of the vector $\overrightarrow{w}_k := \begin{bmatrix}
0 & w_k(1) & \ldots & w_k(N-1) \end{bmatrix}^T$, $k \in \{\mathcal{N}_j \cup j\} \backslash \{i\}$ and $W_\ell^N \in \mathbb{R}^{N\times N}$ the Toeplitz matrix of the vector $\overrightarrow w_\ell := \begin{bmatrix}
0 & w_\ell(1) & \ldots & w_\ell(N-1) \end{bmatrix}^T$ where $\ell \in \{i,j\}$. Similarly, we denote by $\overleftrightarrow W_\ell \in \mathbb{R}^{N\times l}$ the Toeplitz matrix of the vector $\overleftrightarrow{w}_\ell := \begin{bmatrix} 0 & 0 & -w_\ell(1) & \ldots & -w_\ell(N-2) \end{bmatrix}^T$, $\ell \in \{i,j\}$. Also $G_{b}$ and $G_f$ are the Toeplitz matrix of $b_{ji}$ and $f_{ji}$ respectively. Considering the parameterization of $G_{ji}^0$ and the above established notations, we can rewrite the network dynamics in \eqref{eq:43} as \eqref{eq:44} where $\tilde W := W_j + G_{b}\overleftrightarrow W_{i} - G_{f}\overleftrightarrow W_{j}$, $\mathbf W_{ji} = [\begin{matrix} W_{i}^N & -W_{j}^N \end{matrix}]$, $g_{ji} = [\begin{matrix} b_{ji}^\top & f_{ji}^\top \end{matrix}]^\top$ and $\bar e_j$ is the vectorized noise. \hfill \qed
\end{lemma}



\subsection{Modeling strategy for the additional modules}\label{strategy}
We now have a vector description of the module dynamics where we have isolated the objective of the identification method, namely $g_{ji}$, from the non-interesting nuisance terms, namely $m_k$ and $m_j$. As the next step, we discuss our modeling strategy with the use of regularized kernel-based methods.
Our goal is to limit the number of parameters necessary to describe $w_j$ in \eqref{eq:44}, in order to increase the accuracy of the estimated parameter vector of interest $\theta$. In order to achieve this, we keep a parametric model for $g_{ji}$ (accounting for the zeros in $g_{ji}$), while the remaining impulse responses in \eqref{eq:44} are modeled with non-parametric model as zero mean Gaussian processes. The choice of Gaussian processes is motivated by the fact that, with a suitable choice of the prior covariance matrix (usually referred to as kernel), we can get a significant reduction in the variance of the estimated impulse responses \cite{Pillonettoetal_Autom:14}. Therefore, we model $m_j$ and $m_k$, $k \in \mathcal{N}_j \backslash \{i\}$, as independent\footnote{It is clear that these impulse responses share some common dynamics given by the pre-multiplication with the inverse of the noise model $H_j(q)$. However, for computational purposes it is convenient to treat the impulse responses as independent. Furthermore, incorporating the mutual dependence through a suitable choice of prior distribution seems a non-trivial problem that deserves a thorough analysis that is outside the scope of this paper.} zero mean Gaussian processes (vectors in this case). The choice of the covariance matrix (kernel) of these vectors are given by the \textit{First-order Stable Spline kernel} whose general structure is given as,
\begin{equation} \label{eq:ssk}
[{K_\beta}]_{x,y} = \lambda \beta^{\max(x,y)} \,,
\end{equation}
where ${\beta}_j \in [0, 1)$ is a \emph{hyperparameter} that regulates the decay velocity of the realizations of the corresponding Gaussian vector, while $\lambda \geq 0$ tunes their amplitude. The choice of this kernel is motivated by the fact that it enforces favorable properties such as stability and smoothness in the estimated impulse responses \cite{Pillonetto&DeNicolao_Autom:11}, \cite{Pillonetto&DeNicolao_Autom:10}. Therefore, we have that
\begin{align}
{m_{j}} & \sim \mathcal{N}(0, \lambda_j {K_\beta}_j) \label{eq:prior_sj} \\
{m_{k}} & \sim \mathcal{N}(0, \lambda_k {K_\beta}_k) \quad,\, k \in \mathcal{N}_j \backslash \{i \},  \label{eq:prior_sk}
\end{align}
where we have assigned different hyperparameters to the impulse response priors to guarantee flexible enough models.

\subsection{Incorporating Empirical Bayes approach}\label{EB}
We define
\begin{equation}
\bm := \begin{bmatrix}
m_j^\top  & {m_{k}}_1^\top & {m_{k}}_2^\top & \dots & {m_{k}}_p^\top
\end{bmatrix}^\top \,,
\end{equation}
where $k_1,\,\ldots,\,k_p$ are the elements of the set $\mathcal{N}_j \backslash \{i \}$, and
\begin{equation}
\mathbf W := \begin{bmatrix}
\tilde W & {W_{k}}_1 & {W_{k}}_2 & \dots & {W_{k}}_p
\end{bmatrix}\,,
\end{equation}
\begin{equation}
\mathbf K := \mathrm{diag}\lbrace {\lambda_j}{K_{\beta_j}}, {\lambda_k}_1{K_{{\beta_k}_1}}, \dots, {\lambda_k}_p{K_{{\beta_k}_p}}\rbrace.
\end{equation}
Using the above, we can rewrite \eqref{eq:44} in compact form as
\begin{equation}\label{eq:50}
w_j = \mathbf W \bm + \mathbf W_{ji}\bg_{ji} + \bar e_j \,.
\end{equation}

Having assumed a Gaussian distribution of the noise, we can write the joint probabilistic description of $m$ and $w_j$, which is jointly Gaussian, as:
\begin{equation}
p\Bigg(\begin{bmatrix}
\bm \\ w_j
\end{bmatrix}; \eta \Bigg) \sim \mathcal{N}\Bigg(\begin{bmatrix}
\mathbf 0 \\ \mathbf W_{ji}\bg_{ji}
\end{bmatrix}, \begin{bmatrix}
\mathbf K & \mathbf K \mathbf W^\top\\ \mathbf W \mathbf K & \mathbf P
\end{bmatrix}\Bigg),
\end{equation}
where
\begin{equation}
\mathbf P := \bar\sigma_j^2I_N + \tilde W{\lambda_{j}}{{K_\beta}_j}\tilde W + \sum_{ k \in \mathcal{N}_j \backslash \{i \}} {W_{k}}{\lambda_{k}}{{K_\beta}_k}{W_{k}}^\top,
\end{equation}
and this pdf depends upon the vector of parameters
$$\eta := \begin{bmatrix}
\theta^\top &
{\lambda_j} &
{\lambda_k}_1 &
\dots &
{\lambda_k}_p &
{\beta_j} &
{\beta_k}_1 &
\dots &
{\beta_k}_p &
\bar\sigma_j^2
\end{bmatrix} ,$$
which contains the parameter vector of the target module, the hyperparameters of the kernels of the impulse response models of the other modules, and the variance of the ``dummy" noise corrupting $w_j(t)$. Therefore, we focus on the estimation of $\eta$, since it contains the parameter of interest $\theta$. To this end, we apply an Empirical Bayes (EB) approach. We consider the marginal pdf of $w_j$, which is obtained by integrating out the dependence on $\bm$ and corresponds to
\begin{equation}
p(w_j;\eta) \sim \mathcal{N}(\blue{\mathbf W_{ji}}\bg_{ji},\mathbf P).
\end{equation}
Then, the estimate of $\eta$ is obtained by maximizing the marginal likelihood of $w_j$, namely
\begin{equation}\label{eq:2.1}
\begin{split}
&\hat\eta = \argmax_\eta p(w_j; \eta)\\
&=\! \argmin_\eta \log\det\mathbf P \!\!+\!\! \big(w_j - \mathbf W_{ji}\bg_{ji}\big)^\top\!\mathbf P^{-1}\!\big(w_j - \mathbf W_{ji}\bg_{ji}\big).
\end{split}
\end{equation}
Solving this optimization problem can be a cumbersome task, because it is a nonlinear one and involves a large number of decision variables. In the next section, we study how to solve the marginal likelihood problem through a dedicated iterative scheme.

\section{Solution to the marginal likelihood problem}\label{ML}
In this section, we focus on solving the problem in \eqref{eq:2.1} by deriving an iterative solution scheme through the EM algorithm \cite{Dempsteretal_JRSS:1977}.
For this, we need to first define a \emph{latent variable} whose estimation simplifies the computation of the marginal likelihood. In our case, a natural choice is $m$. Then, the solution to \eqref{eq:2.1} using the EM algorithm is obtained by iterating among the following two steps:
\begin{itemize}
	\item \emph{E-Step:} Given an estimate $\hat{\eta}^{(n)}$ computed at the $n^{th}$ iteration, compute
	\begin{equation}\label{eq:4.1}
	Q^{(n)}(\eta) = \mathbb{E}[\log p(w_j,\bm;\eta)] \,,
	\end{equation}
	where the expectation of the joint log-likelihood of $w_j$ and $\bm$ is taken with respect to the posterior $p(\bm|w_j;\hat{\eta}^{(n)})$;
	\item \emph{M-Step:} Update $\hat\eta$ by solving
	\begin{equation} \label{eq:Q_fun}
	\hat{\eta}^{(n+1)} = \argmax_\eta Q^{(n)}(\eta) \,.
	\end{equation}
\end{itemize}
When iterating among the above steps, convergence to a stationary point of the marginal likelihood is ensured \cite{Boyles_JRSS:1983}. \red{This stationary} \blue{point can be a local or global maximum of the objective function}. In the next section, we show that we clearly get an advantage in solving the original marginal likelihood problem \eqref{eq:2.1} by repetitively solving \eqref{eq:Q_fun} using the EM algorithm. We show that, when we use the EM method, the nonlinear optimization problem becomes a problem of iteratively constructing analytical solutions and solving scalar optimization problems, which significantly simplifies solving \eqref{eq:2.1}.


\subsection{Computation of E-step}\label{estep}
First we focus on the E-step. The posterior distribution of $\bm$ given $w_j$ and an estimate of $\eta$ is Gaussian and corresponds to (see also \cite{anderson_moore_1979}),
\begin{equation}\label{eq:6.1}
p(\bm|w_j;\eta) \sim \mathcal{N}\big(\mathbf C(w_j - \mathbf W_{ji}\bg_{ji}), \mathbf P_m\big)
\end{equation}
where
$$
\mathbf P_m = \left(\frac{\mathbf W^\top \mathbf W}{\bar\sigma_j^2} +  \mathbf K^{-1}\right)^{-1}; \quad \mathbf C = \frac{\mathbf P_m \mathbf W^\top}{\bar\sigma_j^2}.
$$

Let $\hat{\bm}^{(n)}$ and $\hat{\mathbf P}_m^{(n)}$ be the posterior mean and covariance of $\bm$ obtained from \eqref{eq:6.1} using $\hat\eta^{(n)}$. We define
$$
\hat{\mathbf M}^{(n)} := \hat{\mathbf P}_m^{(n)} + \hat{\bm}^{(n)}\hat \bm^{(n)\top},
$$
and consider its $l \times l$ diagonal blocks, which we denote by $\hat{\mathbf M}^{(n)}_j$, $\hat{\mathbf M}^{(n)}_{k_1},\,\dots,\,\hat{\mathbf M}^{(n)}_{k_p}$, respectively. These sub-matrices correspond to the posterior second moments of the estimated impulse responses $\hat{m}^{(n)}_j$,$\hat{m}^{(n)}_{k_1},\,\dots,\,\hat{m}^{(n)}_{k_p}$.

The following lemma provides the structure of the function $Q^{(n)}(\eta)$ for the particular situation of our setup in \eqref{eq:2.1}.

\begin{lemma} \label{lemma1}
	Let $\hat\eta^{(n)}$ be the estimate of $\eta$ at the $n^{th}$ iteration of the EM algorithm according to \eqref{eq:Q_fun}. Then
	\begin{equation}\label{eq:Q.1}
	\begin{split}
	Q^{(n)}(\eta) = Q_0^{(n)}(\bar \sigma_j^2,\theta) + \sum_{k \in \{\mathcal{N}_j\cup j\} \backslash \{i \}}  {Q_m}_k^{(n)}({\lambda_k},{\beta_k})
	\end{split}
	\end{equation}
	where
	\begin{equation}\label{eq:14.1}
	\begin{split}
	Q_o^{(n)}(\bar \sigma_j^2,\theta) \!=\! &-N\log(\bar \sigma_j^2) - \frac{1}{\sigma_j^2}\bigg[w_{j}^\top w_{j} - 2w_{j}^\top \mathbf W_{ji}\bg_{ji} + \\
	& \bg_{ji}^\top \mathbf W_{ji}^\top \mathbf W_{ji}\bg_{ji} - 2w_{j}^\top \mathbf W\hat{\bm}^{(n)} \\
	&+ 2\bg_{ji}^\top \mathbf W_{ji}^\top \mathbf W\hat{\bm}^{(n)} + \mathrm{tr}\big(\mathbf W^\top \mathbf W\hat{\mathbf M}^{(n)}\big)\bigg],
	\end{split}
	\end{equation}
	\begin{equation}\label{eq:102}
	\begin{split}
	{Q_m}_k^{(n)}\!({\lambda_k},{\beta_k}) \!=\! &\!-\!\log \det({\lambda_k}{{K_\beta}_k}) \!-\!
	\mathrm{tr}\big({({\lambda_k}{{K_\beta}_k})}^{-1}\hat{\mathbf M}_k^{(n)}\big).
	\end{split}
	\end{equation}
	\hfill$\Box$
\end{lemma}
{\bf Proof}: See the appendix.

The function ${Q}^{(n)}(\eta)$ is the summation of several terms that depend on different components of the vector $\eta$. In particular, we have a term of the type ${Q_m}_k^{(n)}({\lambda_k},{\beta_k})$ for each module in the MISO structure, and a term $Q_0^{(n)}(\bar\sigma_j^2,\theta)$ for the module of interest and the noise variance. Therefore, the update of $\eta$ according to \eqref{eq:Q_fun} splits into a number of independent and smaller optimization problems.

\subsection{Computation of M-step}\label{mstep}
We now focus on the M-step according to \eqref{eq:Q_fun}. From \eqref{eq:Q.1}, it is evident that each kernel hyperparameters can be updated independently of the rest of the parameters.  The following theorem, inspired by \cite{Bottegaletal_Autom:16} and \cite{Everitt&Bottegal&Hjalmarsson_Autom:18}, shows how to update the kernel hyperparameters.
\begin{theorem}\label{theorem1}
	For the update of each kernel's hyperparameters that requires maximizing \eqref{eq:102}, we define
	\begin{equation}\label{eq:101}
	{Q_\beta}_k^{(n)}({\beta_k}) = \log \det({{K_\beta}_k}) + l \log \bigg(\mathrm{tr}\big({({{K_\beta}_k})}^{-1}\hat{\mathbf M}_k^{(n)}\big)\bigg)
	\end{equation}
	for  $k \in \{\mathcal N_j\cup j\} \backslash i$.
	Then the updates are obtained as,
	\begin{equation}\label{eq:12.1}
	{\hat{\beta_k}}^{(n+1)} = \argmin_{{\beta_k} \in [0,1)}{Q_\beta}_k^{(n)}({\beta_k}); \end{equation}
	\begin{equation}\label{eq:13.1}
	{\hat{\lambda_k}}^{(n+1)} = \frac{1}{l}\mathrm{tr}\big({(K_{{\hat\beta}_{{k}}^{(n+1)}})}^{-1}\hat{\mathbf M}_k^{(n)}\big)
	\end{equation}
	\hfill$\Box$
\end{theorem}
{\bf Proof}: See the appendix.

\blue{The optimization problem in \eqref{eq:12.1} can be difficult to perform in practice when the determinant of the kernel has a very low value or when the inversion of the kernel becomes difficult. To tackle this, we exploit the factorization of the \emph{first order stable spline kernel} as in \cite{Bottegaletal_Autom:16} by writing ${K_\beta}_k = L D(\beta) L^T$, where $L$ is lower-triangular with known entries (essentially, an ``integrator'') and $D(\beta)$ is diagonal with entries essentially being an exponential functions of $\beta$. Using the above technique also increases the computation speed of the algorithm.}

We note that from \eqref{eq:13.1} that we get closed-form solutions for all ${\lambda_k}$, $k \in \{\mathcal N_j\cup j\} \backslash \{i\}$, while the ${\beta_k}$, $k \in \{\mathcal N_j\cup j\} \backslash \{i\}$, can be updated by solving scalar optimization problems in the domain $[0,1)$, as detailed in \eqref{eq:12.1}. Therefore, the hyperparameters update turns out to be a computationally fast operation.

We now turn our attention to the update of $\theta$ and $\bar \sigma_j^2$ for which we need to maximize \eqref{eq:14.1}.
We notice that the optimum with respect to $\theta$ does not depend on the optimal value of $\bar \sigma_j^2$. Then, we can first update $\theta$ and then use its optimal value to update $\bar \sigma_j^2$. How to update $\theta$ is explained in the following theorem.

\begin{theorem}\label{theorem2}
	The estimate of the parameter vector $\theta$ is updated by solving the quadratic problem
	\begin{equation}
	\hat{\theta}^{(n+1)} = \argmin_\theta \bigg[\bg_{ji}^\top \hat{\mathbf A}^{(n)}\bg_{ji} - 2\hat{\mathbf{b}}^{(n)\top}\bg_{ji}\bigg]
	\end{equation}
	that has a closed form solution given by
	\begin{equation}\label{eq:54}
	\hat{\theta}^{(n+1)} = \big(M^\top \hat{\mathbf A}^{(n)}M\big)^{-1}M^\top\hat{\mathbf b}^{(n)},
	\end{equation}
	where $\hat{\mathbf A}^{(n)}$ and $\hat{\mathbf b}^{(n)}$ are computed using the current estimates $\hat{\bm}^{(n)}$ and $\hat\eta^{(n)}$, and $\bg_{ji} = M\theta$ where $M \in \mathbb{R}^{2N\times n_{\theta}}$ is a matrix with 1 or 0 as its elements. \hfill$\Box$
\end{theorem}
{\bf Proof}: See the appendix.

Therefore, the parameter vector of the target module is updated by solving the analytical expression \eqref{eq:54}.
\begin{remark}
    An additional advantage of the method developed in this paper is that it relies on iteratively solving a quadratic least squares problem to find the solution for the parameters of the target module $\theta$ rather than solving a non-linear least squares problem as in \cite{Ramaswamy&etal_CDC:18}, making the method computationally more efficient.
\end{remark}
We are left with updating $\bar \sigma_j^2$, which is given in the next theorem.
\begin{theorem} \label{theorem3}
	Let $\hat{\bg}_{ji}^{(n+1)}$, $\mathbf{\hat{W}}^{(n+1)}$ be constructed by inserting $\hat{\theta}^{(n+1)}$ in the general expression of $\bg_{ji}$ and $\mathbf W$. Then
	\begin{equation*}\label{eq:55}
	\begin{split}
	(&\hat{\bar \sigma}_j^2)^{(n+1)} \!=\! \frac{1}{N}\bigg[{\|w_j - \mathbf W_{ji}\hat{\bg}_{ji}^{(n+1)}\|}_2^2 - 2w_{j}^\top \mathbf{\hat{W}}^{(n+1)}\hat{\bm}^{(n)} +\\ &2\hat{\bg}_{ji}^{(n+1)\top}\mathbf W_{ji}^\top \mathbf{\hat{W}}^{(n+1)}\hat{\bm}^{(n)} \!+\! \mathrm{tr}\big(\mathbf{\hat{W}}^{(n+1)\top} \mathbf{\hat{W}}^{(n+1)}\hat{\mathbf M}^{(n)}\big) \bigg]
	\end{split}
	\end{equation*}
	\hfill$\Box$
\end{theorem}
{\bf Proof}: See the appendix.

Thus, a closed-form solution for the estimate of the noise variance is also obtained.
\begin{remark}\label{rem:noise}
	We estimate the ``dummy" noise variance $\bar \sigma_j^2 = {|{f_{a}}_{n_f}|}^2\sigma_j^2$, that is a scaled version of the original output noise power in the network. If there are no unstable systems in the MISO setup, then $\bar \sigma_j^2$ will be $\sigma_j^2$. This will be verified with numerical simulations in section \ref{sec:Num_ex}.
\end{remark}

All-in-all, we have obtained a fast iterative procedure that provides a local solution to the marginal likelihood problem \eqref{eq:2.1}. All the updates follow simple rules that allow for fast iterative computation. Algorithm 1 summarizes the steps to follow to obtain $\hat \eta$ and therefore $\hat \theta$.
\begin{algorithm}[h]\label{algo:3}
	\textbf{Input:} $\{w_k(t)\}_{t=1}^N$, $k =1,\ldots,p$ \\
	\textbf{Output:} $\hat \theta$
	\begin{enumerate}
		\item Set $n = 0$, Initialize $\hat{\eta}^{(0)}$.
		\item Compute $\hat{\mathbf P}_m^{(n)}$, $\hat{\mathbf C}^{(n)}$, $\hat{\mathbf M}^{(n)}$ and $\hat{\bm}^{(n)}$.
		\item Update hyperparameters ${\hat{\beta_k}}^{(n+1)}$ and ${\hat{\lambda_k}}^{(n+1)}$ using \eqref{eq:12.1} and \eqref{eq:13.1} respectively for all $k \in \{\mathcal{N}_j\cup \{j\}\}\backslash\{i\}$.
		\item Update $\hat{\theta}^{(n+1)}$ by solving \eqref{eq:54}.
		\item Update $\hat{\bar \sigma}_j^{2(n+1)}$ as in Theorem \ref{theorem3}.
		\item Set $\hat{\eta}^{(n+1)}\\ = [\begin{smallmatrix}
		\hat{\theta}^{\top(n+1)} &
		{\hat{\lambda_j}}^{(n+1)}  &
		{\hat{\lambda_k}}_1^{(n+1)}  &
		\dots &
		{\hat{\lambda_k}}_p^{(n+1)}
		\end{smallmatrix}\\
		\begin{smallmatrix}
		\quad \quad \quad \quad \quad \quad \quad \quad {\hat{\beta_j}}^{(n+1)}  &
		{\hat{\beta_k}}_1^{(n+1)}  &
		\dots &
		{\hat{\beta_k}}_p^{(n+1)}  &
		{\hat{\bar\sigma}}_j^{2(n+1)}
		\end{smallmatrix}]^\top$
		\item Set $n = n + 1$.
		\item Repeat from steps (2) to (7) until convergence.
	\end{enumerate}
	\caption{Algorithm for local module identification in dynamic networks}
\end{algorithm}

The initialization can be done by randomly choosing $\eta$ considering the constraints of hyperparameters. The convergence criterion for the algorithm depend on the value of $\frac{\norm{\hat{\eta}^{(n)} - \hat{\eta}^{(n-1)}}}{\norm{\hat{\eta}^{(n-1)}}}$. This value should be small for convergence so that the algorithm can be terminated. A value of $10^{-2}$ is considered for the numerical simulations in Section \ref{sec:Num_ex}. The other convergence criterion is the maximum number of iterations. It is taken as 50.

\begin{remark}
	Being applicable to a MISO identification setup, the introduced method can also be inherently used for parametric SISO identification, where the process noise modeling is now simplified by avoiding the model order selection and reducing the number of parameters of the noise model to two (which are the hyperparamters $\lambda_{j}, \beta_j$).
\end{remark}

\begin{remark}
	We notice that:
	\begin{itemize}
	    \item The method does not require prior information about the stability of the systems $G_{jk}, k \in \mN_j$ and the number of unstable poles in the systems. 
	    \item According to \cite{Dankers&etal_TAC:16}, \blue{in view of consistency of the target module estimate,} it is not necessary to take all nodes $w_k, k \in \mN_j$ as the inputs in the MISO structure, but it is sufficient to take a subset of nodes in $\mN_j$ as inputs such that every parallel path\footnote{a path from $w_i$ to $w_j$ that does not pass through $G_{ji}$.} from $w_i$ to $w_j$ and every loop around $w_j$ passes through a selected input. This may lead to confounding variables which can be handled using additional inputs\cite{Dankers&etal_IFAC:17}. \blue{At the same time, in view of an appropriate bias-variance trade off, especially under limited data circumstances, it could be attractive to include more predictor inputs than the ones that are strictly necessary for achieving consistency. While the algorithm presented in this paper can be applied to any choice of such MISO structure, we have formulated the results for the situation where all nodes $w_k, k \in \mathcal{N}_j$ are taken as inputs. }
	\end{itemize}
\end{remark}

\subsection{Non-parametric identification of modules in the MISO structure}
In this section we slightly adapt the developed method to obtain a non-parametric estimate of the target module. For this, we rewrite the network equation \eqref{eq:singleblock} as,
	\begin{align} \label{eq:351}
	w_j(t) &= {M_j(q)}w_j(t) + \sum_{k \in \mathcal{N}_j} M_{jk}(q)w_k(t) + \bar e_j(t)
	\end{align}
	with
	\beqr
	{M_j(q)} &:=& 1 - \bigg((H_j)^{-1}(q)\frac{F_a(q)}{F_a^\star(q)}\bigg) \quad,\\ M_{jk}(q) &:=& (H_j)^{-1}\frac{\prod_{\ell\in \mN_j \backslash \{ k\}}F_{j\ell}^{(a)}(q)}{F_a^\star(q)}\frac{B_{jk}(q)}{F_{jk}^{(s)}(q)} \, ,
	\eeqr
where $M_{jk}(q)$ and $M_{j}(q)$ are stable. Following the similar approach as introduced before, but modeling the impulse response of all the modules (including $m_i$ of $M_{ji}$ that represents the target module) as zero mean Gaussian processes with the prior covariance matrix represented by the First-order stable spline kernel, we end up in an iterative algorithm to estimate the parameter vector $\eta$ which contains the hyperparameters $\lambda_k, \beta_k$ where $k \in \mN_j$ and the noise variance $\bar \sigma_j^2$. Since we are not paramterizing any modules, we do not have $\theta$ in the parameter vector $\eta$. The solutions for the $\beta$'s and $\lambda$'s at each iteration are given by \eqref{eq:12.1} and \eqref{eq:13.1} respectively. The solution to $\bar \sigma_j^2$ at each iteration is given by,
\begin{equation*}\label{eq:17.2}
\begin{split}
(&\hat{\bar \sigma}_j^2)^{(n+1)} = \frac{1}{N}\bigg[{\|w_j\|}_2^2 - 2w_{j}^\top \mathbf{{W}}\hat{\bm}^{(n)} + \mathrm{tr}\big(\mathbf{{W}}^{\top} \mathbf{{W}}\hat{\mathbf M}^{(n)}\big) \bigg]
\end{split}
\end{equation*}
where \begin{equation*}
\mathbf W := \begin{bmatrix}
W_j & {W_{k}}_1 & {W_{k}}_2 & \dots & {W_{k}}_p
\end{bmatrix}.
\end{equation*}
The above solution is equivalent to the solution of $\hat{\bar \sigma}_j^2$ in Theorem \ref{theorem3}, however without the terms that are function of $\theta$ (i.e. $\bg_{ji}, G_{b}, G_f, \mathbf W_{ji}\bg_{ji}$). Thus we will end up in the same Algorithm 1, however with steps related to $\theta$ (step 4) being not applicable. The posterior mean of $m_k, k \in \mN_j$ and $m_j$ obtained using \eqref{eq:6.1} (neglecting the effect of $W_{ji}g_{ji}$) for the converged $\eta$ provides us the impulse response of $M_{jk}$ and $M_j$ respectively. From these, the impulse response estimates of the modules $G_{jk}, k \in \mN_j$ can be obtained. Thus we obtain a non-parametric identification method to identify all the modules in the MISO structure as a derived result of the earlier developed identification technique.

\section{Numerical simulations}\label{sec:Num_ex}
Numerical simulations are performed to evaluate the performance of the developed method, which we abbreviate as Empirical Bayes Direct Method (EBDM). The simulations are performed on the dynamic network depicted in Figure \ref{fig:dynnet_Ex_wnoise1}. The goal is to identify $G_{31}^0$. To show the effectiveness of the introduced method and its flexibility to handle stable and unstable modules with a single unified identification framework, we perform the simulations for two different cases:
\begin{enumerate}
	\item Case 1: All modules in the MISO setup are stable.
	\item Case 2: The modules in the MISO setup including the target module can be stable or unstable.
\end{enumerate}
The results of the numerical simulations are presented below.

\subsection{Case study 1}
The EBDM is compared with the standard direct method and the two-stage method (see \cite{VandenHof&etal_Autom:13} for details).
The network modules of network in Figure \ref{fig:dynnet_Ex_wnoise1} are given by
\begin{align*}
&G_{31}^0 = \frac{q^{-1} + 0.05q^{-2}}{1 + q^{-1} + 0.6q^{-2}} = \frac{b_1^0q^{-1} + b_2^0q^{-2}}{1 + a_1^0q^{-1} + a_2^0q^{-2}}\\
&G_{32}^0 = \frac{0.09 q^{-1}}{1 + 0.5 q^{-1}};\\
&G_{34}^0 = \frac{1.184 q^{-1} - 0.647 q^{-2} + 0.151 q^{-3} - 0.082 q^{-4}}{1 - 0.8 q^{-1} + 0.279 q^{-2} - 0.048 q^{-3} + 0.01 q^{-4}};\\
&G_{14}^0 = G_{21}^0 = \frac{0.4q^{-1} - 0.5q^{-2}}{1 + 0.3q^{-1}};H_{1}^0 = \frac{1}{1 + 0.2q^{-1}};\\
&G_{12}^0 = G_{23}^0 = \frac{0.4q^{-1} + 0.5q^{-2}}{1 + 0.3q^{-1}};H_{2}^0 = \frac{1}{1 + 0.3q^{-1}}\\
&H_{3}^0 = \frac{1 - 0.505 q^{-1} + 0.155 q^{-2} - 0.01 q^{-3}}{1 - 0.729 q^{-1} + 0.236 q^{-2} - 0.019 q^{-3} }; H_{4}^0 = 1.
\end{align*}
We run $50$ independent Monte Carlo experiments where the data is generated using known reference signals $r_2(t)$ and $r_4(t)$ that are realizations of white noise with unit variance. The number of data samples is $N$ = 500. The noise sources $e_1(t)$, $e_2(t)$, $e_3(t)$ and $e_4(t)$ have variance 0.05, 0.08, 0.5, 0.1, respectively. We assume that we know the model order of $G_{31}^0(q)$. In the case of direct method, we solve a 3-input/1-output MISO identification problem with $w_1(t)$, $w_2(t)$ and $w_4(t)$ as inputs. In the two-stage method, the projections of the three inputs on external signals $r_2(t)$ and $r_4(t)$ are used as inputs to the MISO identification problem. For both these methods, we consider the case where a model order selection of all the modules in the MISO structure (except for the target module) is required, and the case where the model orders are known. Moreover, in order to improve the accuracy of the identified module in the two-stage method, we identify a noise model even though it is not necessary for consistency.

\begin{figure}
	\centering
	\includegraphics[scale=0.22]{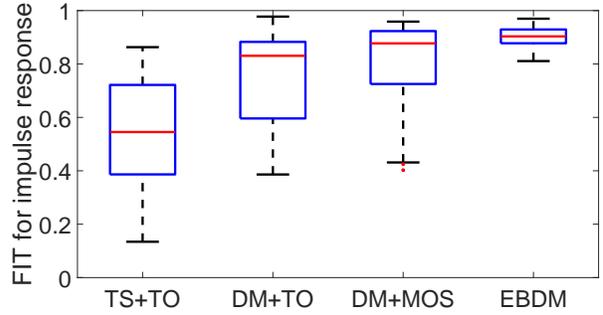}
	\caption{Box plot of the fit of the impulse response of $\hat{G}_{31}$ obtained by the Two-stage method, Direct method and EBDM. Number of data samples used for estimation is $N$ = 500.}
	\label{fig:boxplotimp}
\end{figure}

\begin{figure}
	\centering
	\includegraphics[scale=0.22]{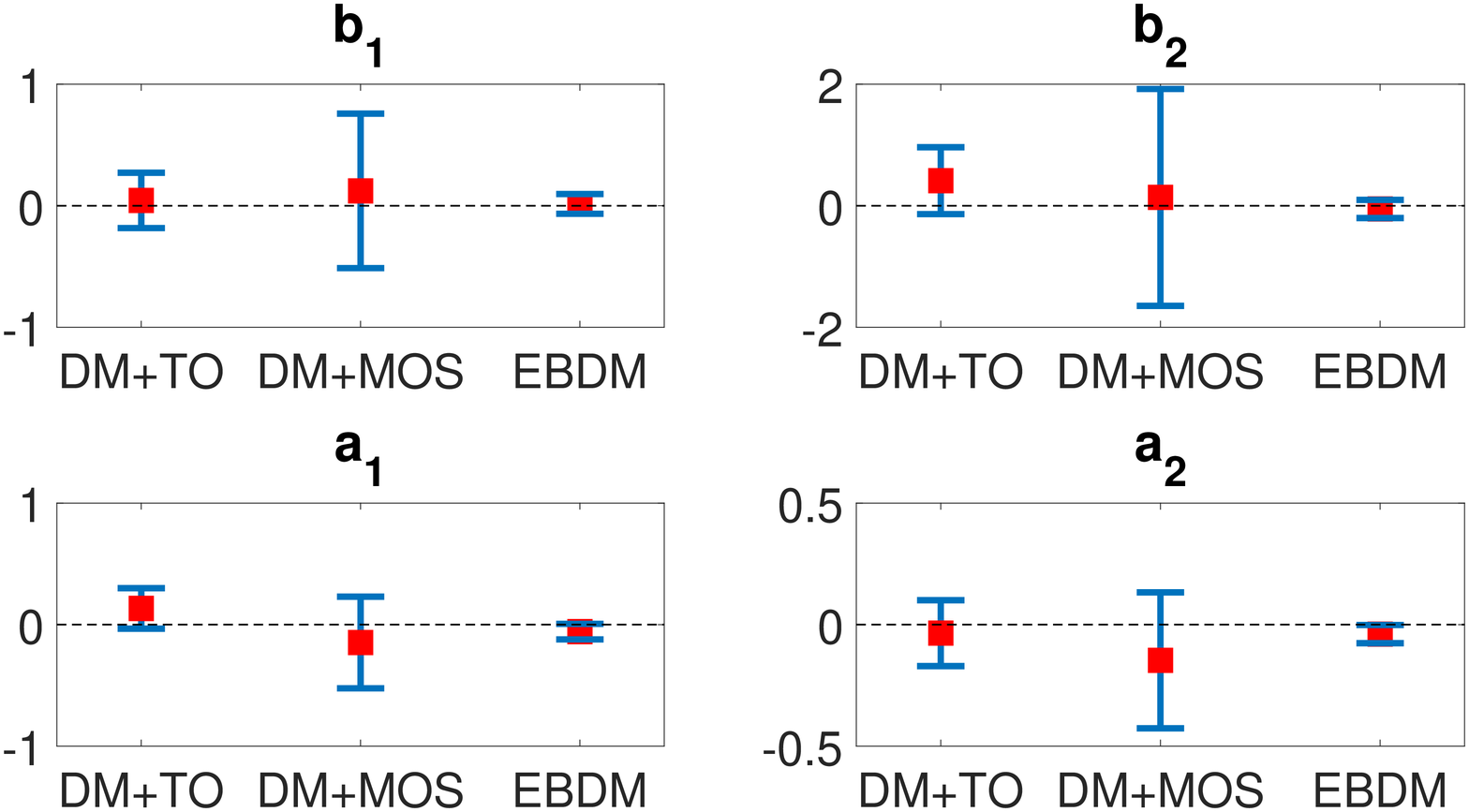}
	\caption{Bias and standard deviation of each parameter obtained from 50 MC simulations using different identification methods.}
	\label{fig:errorplotmse}
\end{figure}

\begin{figure}
\begin{flushleft}
	\includegraphics[trim={4.15cm 0 0 0},clip,scale=0.25]{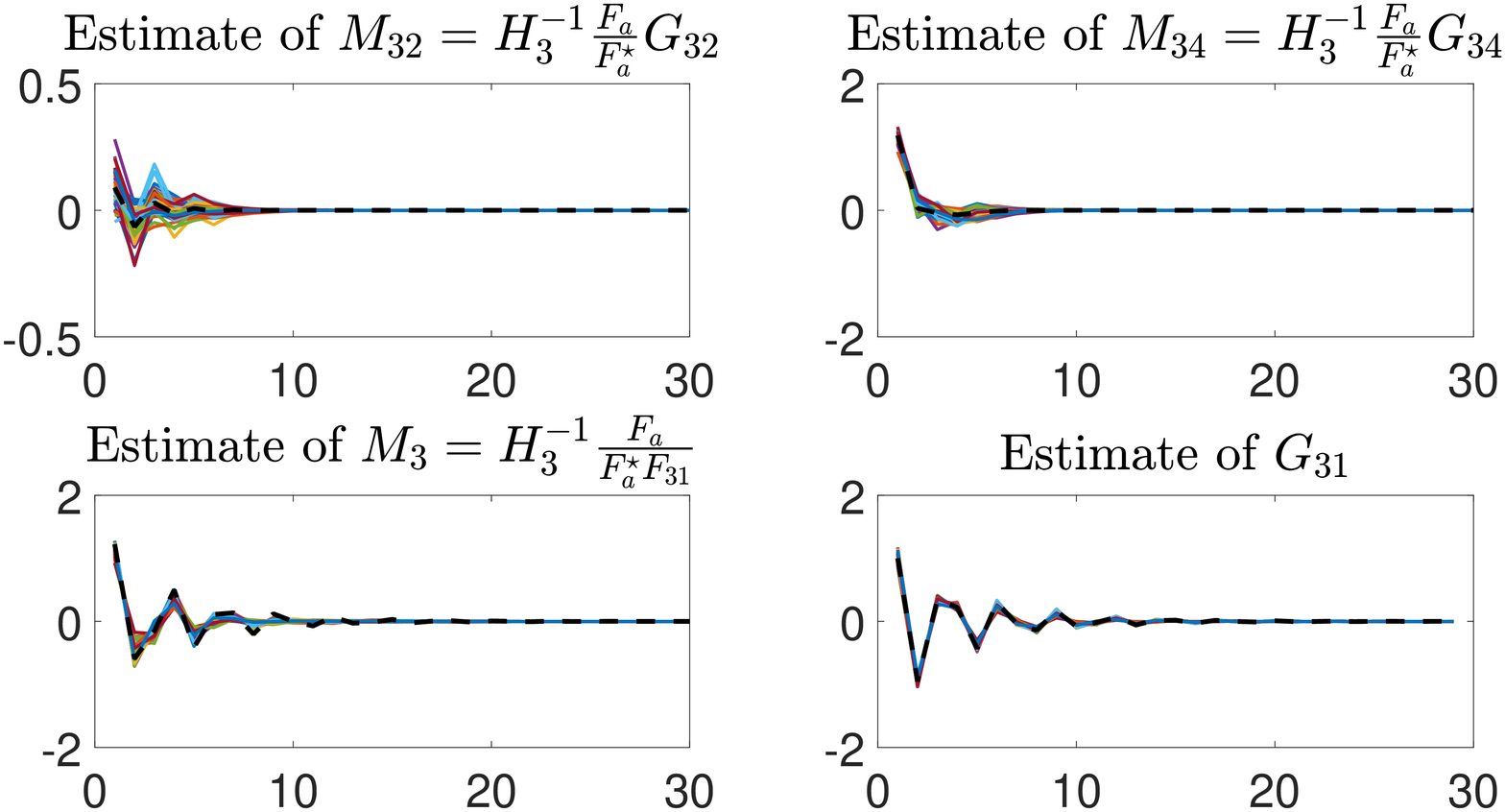}
	\caption{Bottom right plot provides the impulse response estimate of the target module at the end of each MC simulation, which is obtained from the estimated parameter $\theta$. The other plots show the impulse response estimates of the filters that are modeled as GP's, which is obtained by calculating the posterior \eqref{eq:6.1} from the estimated hyperparameters. The black dashed line provides the true impulse response of the modules.}
	\label{fig:posterior}
\end{flushleft}
\end{figure}

Figure \ref{fig:posterior} shows the estimated impulse response at the end of each MC simulation using the EBDM. It can be verified that, in line with our framework, the estimates provide the description of the dynamics of $M_j$, $M_{jk}, k \in \mN_j$ and $G_{ji}$. To evaluate the performance of the methods, we use the standard goodness-of-fit metric,
\beq
\textrm{Fit} = 1 - \frac{{\norm{g_{ji}^0 - \hat g_{ji}}}_2}{{\norm{g_{ji}^0 - \bar g_{ji}}}_2}, \nonumber
\eeq
where $g_{ji}^0$ is the true value of the impulse response of $G_{ji}^0$, $\hat g_{ji}$ is the impulse response of the estimated target module and $\bar g_{ji}$ is the sample mean of $g_{ji}^0$.
The box plots of the fits of the impulse response of $G_{31}(q)$ are shown in Figure \ref{fig:boxplotimp}, where we have compared the two-stage method with true model orders ('TS+TO'), the direct method with true model orders and model orders selected via BIC ('DM+TO' and 'DM+MOS', respectively), and the Empirical Bayes Direct Method ('EBDM'). As for the latter, we choose $l = 100$. It can be noted that in this setup the EBDM achieves a fit on par with the Direct method and significantly better than the two-stage method. Figure \ref{fig:errorplotmse} shows the mean and standard deviation of the parameter estimates of $G_{31}$. It is evident that the EBDM gives a smaller bias and a greatly reduced variance compared to the other considered identification methods. The reduction in variance is attributed to the regularization approach used in this method. The fit is calculated using the estimated impulse response from the estimated parameters of the target module. Even though, the variability is high in estimated parameters using the Direct Method, it did not affect the fit of the impulse response, that produces an on par result in figure \ref{fig:boxplotimp} when compared with EBDM. However, Figure \ref{fig:errorplotmse} clearly shows that EBDM performs better than the other considered approaches. Considering a relatively small sized network with 3 modules in the MISO structure, the developed method proves effective. When the size of the network grows, the results of the direct method may deteriorate further due to increase in variance; furthermore, it is expected that in large networks the model order selection step contributes to inaccurate results. Thus the EBDM, by offering reduced variance and circumventing the problem of model order selection, can stand out as an effective local module identification method in large dynamic networks.


\subsection{Case study 2}
Now we look into the case where the modules in the MISO structure may not be stable. In this case, we consider the same network as in Figure \ref{fig:dynnet_Ex_wnoise1}, however with unstable module $G_{31}^0$ (target module) and $G_{32}^0$. The network modules of network in Figure \ref{fig:dynnet_Ex_wnoise1} are the same as in previous section but with unstable $G_{31}^0$ and $G_{32}^0$ given by
\begin{align*}
&G_{31}^0 \!\!=\!\! \frac{q^{-1} + 0.05q^{-2}}{1 + 1.7q^{-1} + 1.073q^{-2}} = \frac{b_1^0q^{-1} + b_2^0q^{-2}}{1 + a_1^0q^{-1} + a_2^0q^{-2}}\\
&G_{32}^0 \!\!=\!\! \frac{-0.7339 q^{-1} \!-\! 0.1256 q^{-2} \!+\! 0.04023 q^{-3} \!+\! 0.011 q^{-4}}{1 \!-\! 1.089 q^{-1} \!-\! 0.104 q^{-2} \!+\! 0.052 q^{-3} \!+\! 0.011 q^{-4}}.
\end{align*}
$G_{31}^0$ has two complex poles that are not stable and $G_{32}^0$ has four poles of which one is a real unstable pole. The noise source $e_3(t)$ has variance of 0.1. The experiment setup is similar to the previous case and we run 50 MC experiments with the introduced method in this paper.

\begin{figure}
	\centering
	\includegraphics[scale=0.22]{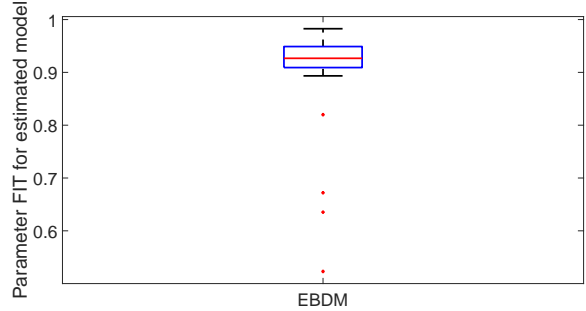}
	\caption{Box plot of the fit of the parameters of $\hat{G}_{31}$ obtained by the proposed method. Number of data samples used for estimation is $N$ = 500.}
	\label{fig:boxplotimp1}
\end{figure}

\begin{figure}
	\centering
	\includegraphics[scale=0.22]{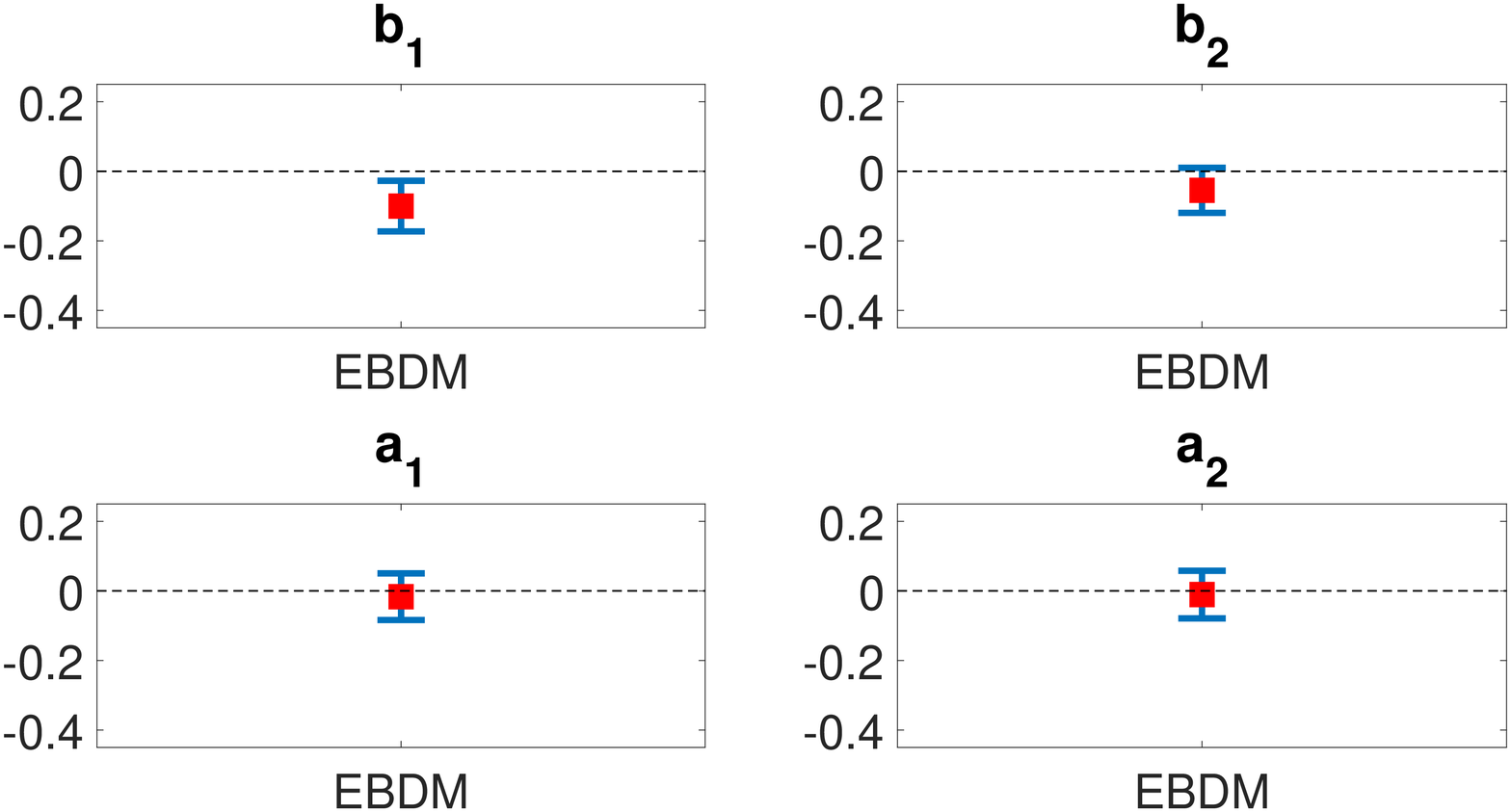}
	\caption{Bias and standard deviation of each parameter obtained from 50 MC simulations using different identification methods.}
	\label{fig:errorplotmse1}
\end{figure}

To evaluate the performance of the EBDM, we use the standard goodness-of-fit metric,
\beq
\textrm{Fit} = 1 - \frac{{\norm{\theta^0 - \hat \theta}}_2}{{\norm{\theta^0 - \bar \theta}}_2}, \nonumber
\eeq
where $\theta^0$ are the true parameters of the target module, $\hat \theta$ are the estimated parameters and $\bar \theta$ is the sample mean of $\theta^0$. Due to the instability of the target module, we choose fit on parameters and not on the impulse response. The box plot of the fit of the parameters of $G_{31}(q)$ is shown in Figure \ref{fig:boxplotimp1}, where the Empirical Bayes Direct Method ('EBDM') is used to identify the unstable target module. We choose $l = 200$. It can be noted that the box plot is above 0.9, which indicates a better fit. Figure \ref{fig:errorplotmse1} shows the mean and standard deviation of the parameter estimates of $G_{31}$. It is evident that the bias and variance is small. The reduction in variance is attributed to the regularization approach used in this method. 

\begin{figure}
	\centering
	\includegraphics[scale=0.22]{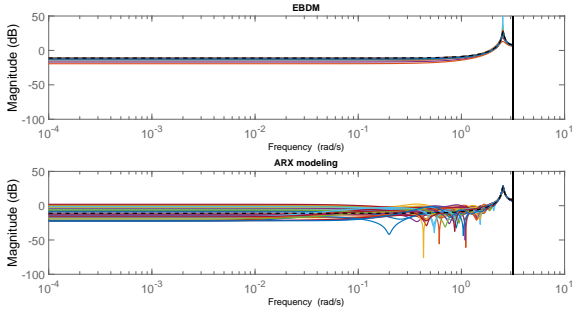}
	\caption{Bode magnitude plot to compare the estimates of the introduced approach(upper) and the approach in \cite{Galrinho&etal_Autom:17}(lower).}
	\label{fig:bodemag}
\end{figure}

It is noteworthy to compare the introduced EBDM with other available approaches that can identify unstable modules. In \cite{Galrinho&etal_Autom:17}, a method to identify unstable SISO systems with Box-Jenkins (BJ) structure using high order ARX modeling has been introduced. This method proves effective in estimating the unstable poles of the system with high accuracy (less variance) \cite{Galrinho&etal_Autom:17}, but the estimated model will have high variance due to high order modeling. Also, the estimated model will be of high order unless there is sufficiently large data. Figure \ref{fig:bodemag} shows the bode magnitude plot of the estimates after 50 MC simulations with the experimental setup in case study 2 using EBDM and the method of ARX modeling in \cite{Galrinho&etal_Autom:17}. ARX models of 15$^{th}$ order are used for the latter method. Even though the estimate of unstable poles are with high accuracy for the latter method, the EBDM performs significantly better in terms of accuracy with less variance in the identified frequency response. Since we have limited data ($N = 500$), the estimated model with the method in \cite{Galrinho&etal_Autom:17} is of high order, which can be verified from figure \ref{fig:bodemag}. 

A three step parametric identification method to identify unstable SISO system is introduced in \cite{Miguel&etal_CDC:2016}. The first step involves identifying the unstable poles of the parameterized model using the result that the unstable poles can be identified with high accuracy using the method in \cite{Galrinho&etal_Autom:17}. In the next step,  from the obtained estimates, the parameters of the anti-stable part is fixed, and a weighted null space fitting (WNSF) method is used to identify the rest of the parameters of the parameterized model of interest. However, for the MISO identification setup in a dynamic network framework, we might end up in estimating 'false' unstable poles for the target module in the first step where ARX modeling is used. Due to high order ARX modeling, these 'false' unstable poles can be the unstable poles of the modules in the MISO setup other than the target module and it becomes difficult to distinguish the unstable poles between each modules, so that the estimate of unstable roots of the target module can be fixed for the second step. For example, the simulations depicted in Figure \ref{fig:bodemag} using the ARX modeling method, we estimate the target module of order 15 with 3 unstable poles, where 2 unstable poles are the poles of $G_{31}^0$ and the extra unstable pole is the unstable pole of $G_{32}^0$. Therefore, it becomes difficult to use the WNSF method in this setup without prior knowledge about the unstable poles. An alternative BJ model has been proposed in \cite{Forssell&Ljung_CDC:1998} that can be used with prediction error framework. However, implementation of this is significantly more complex than the introduced EBDM.


\begin{table*}[]
\begin{tabular}{lllllllllll}
\hline
\multicolumn{1}{|l|}{Actual value ($\bar \sigma_3 = \sigma_3$)} &
  \multicolumn{1}{l|}{0.1} &
  \multicolumn{1}{l|}{0.2} &
  \multicolumn{1}{l|}{0.3} &
  \multicolumn{1}{l|}{0.4} &
  \multicolumn{1}{l|}{0.5} &
  \multicolumn{1}{l|}{0.6} &
  \multicolumn{1}{l|}{0.7} &
  \multicolumn{1}{l|}{0.8} &
  \multicolumn{1}{l|}{1} &
  \multicolumn{1}{l|}{2} \\ \hline
\multicolumn{1}{|l|}{Estimated value ($\hat{\bar \sigma}_3$)} &
  \multicolumn{1}{l|}{0.0971} &
  \multicolumn{1}{l|}{0.1908} &
  \multicolumn{1}{l|}{0.2804} &
  \multicolumn{1}{l|}{0.4093} &
  \multicolumn{1}{l|}{0.4710} &
  \multicolumn{1}{l|}{0.6314} &
  \multicolumn{1}{l|}{0.7620} &
  \multicolumn{1}{l|}{0.8207} &
  \multicolumn{1}{l|}{0.9449} &
  \multicolumn{1}{l|}{1.9398} \\ \hline
 &
   &
   &
   &
   &
   &
   &
   &
   &
   &
   \\ \hline
\multicolumn{1}{|l|}{Actual value ($\bar \sigma_3^2 = {|\frac{F_a}{F_a^\star}|}^2 \sigma_3^2$)} &
  \multicolumn{1}{l|}{0.1475} &
  \multicolumn{1}{l|}{0.2950} &
  \multicolumn{1}{l|}{0.4425} &
  \multicolumn{1}{l|}{0.5901} &
  \multicolumn{1}{l|}{0.7376} &
  \multicolumn{1}{l|}{0.8851} &
  \multicolumn{1}{l|}{1.0326} &
  \multicolumn{1}{l|}{1.1801} &
  \multicolumn{1}{l|}{1.4752} &
  \multicolumn{1}{l|}{2.9503} \\ \hline
\multicolumn{1}{|l|}{Estimated value ($\hat{\bar \sigma}_3$)} &
  \multicolumn{1}{l|}{0.1520} &
  \multicolumn{1}{l|}{0.3005} &
  \multicolumn{1}{l|}{0.4579} &
  \multicolumn{1}{l|}{0.5946} &
  \multicolumn{1}{l|}{0.7338} &
  \multicolumn{1}{l|}{0.8642} &
  \multicolumn{1}{l|}{0.9145} &
  \multicolumn{1}{l|}{1.1851} &
  \multicolumn{1}{l|}{1.6030} &
  \multicolumn{1}{l|}{2.7349} \\ \hline
\end{tabular}
\caption{Results of the simulations that were performed using the setup of case study 1 (upper) and 2 (lower) with different noise variance of $e_3$ acting on the output node $w_3$. Table 1 shows the actual ``dummy" noise variance to be estimated and the estimated noise variance using EBDM for the experimental setup in Case 1 (upper) and Case 2 (lower).}
\label{table:noisevariance}
\end{table*}

\subsection{Estimated noise variance}
Using the experimental setup of case study 1 and 2 but with different noise power (variance) of $e_3$ ($\sigma_3$) acting on the output node $w_3$, we performed simulations using the EBDM for the network in Figure \ref{fig:dynnet_Ex_wnoise1}. For the case study 1, since all modules are stable (i.e. $\frac{F_a}{F_a^\star} = 1$), the estimated noise variance $\hat{\bar\sigma}_3$ should be approximately equal to the actual noise variance $\sigma_3$ (see remark \ref{rem:noise}). This can be verified from the Table \ref{table:noisevariance} (upper) where the estimated noise variance approximates well the actual noise variance in the network. Considering the case study 2, the estimated noise variance $\hat{\bar\sigma}_3$ should be approximately equal to the scaled version of the actual noise variance $\sigma_3$ given by $\bar \sigma_3^2 = {|\frac{F_a}{F_a^\star}|}^2 \sigma_3^2= {|{f_{a}}_{n_f}|}^2\sigma_3^2$ i.e. the ``dummy"noise variance. This can be verified from the Table \ref{table:noisevariance} (lower).

\blue{
\subsection{Additional remarks}
    The method described in this paper can be developed using any of the kernels available in the literature of regularized system identification. The choice of kernel adopted in this paper is the result of a balance between its empirical effectiveness (see \cite{Pillonettoetal_Autom:14}) and its computational efficiency (due to its factorization and the low number of hyperparameter). 
Other choices of kernel (e.g. the DC kernel proposed in \cite{Chenetal_Autom:12}) may result in a final higher accuracy, requiring to estimate an additional hyperparameter, which might bring an additional cost in complexity. On the other hand, it is well known (see \cite{Chenetal_Autom:12}) that the optimal kernel is constructed from the true impulse response, which is unknown (it is the actual object of interest). The question \red{which is} the best choice of kernel for dynamic networks is open and requires a thorough theoretical analysis which is outside the scope of the paper.
}

\section{Conclusions}\label{sec:conclusion}
An effective regularized kernel-based approach for local module identification in dynamic networks has been introduced in this paper. The introduced method (EBDM) circumvents the model order selection step for all the modules that are not of primary interest to the experimenter, but still need to be identified in order to get a consistent estimate of the target module. Furthermore, by using regularized non-parametric methods, the number of parameters to be estimated is greatly reduced, with a clear benefit in terms of mean square error of the estimated target module. Therefore, the method is computationally less complex and scales favorably to large size networks. 
The method developed in this paper is capable of performing identification in networks composed by unstable modules, without any prior information about the stability of the modules.
Numerical experiments performed with a dynamic network example illustrate the potentials of the developed method on comparison with the already available methods on networks of stable modules. The developed method provides better estimates and a reduced variance is observed in the identified model due to the integration of the regularization approach in the method.

\bibliographystyle{plain}        
\bibliography{Paul_Dynamic_Networks_Library}           



\appendix
\section{Proof of Proposition \ref{propnew1}} 
Analogous to the factorization technique used in \cite{Forssell&Ljung_CDC:1998} and \cite{Galrinho&etal_Autom:17}, 
we factorize each $F_{jk}$ (from now on superscript $^0$ is dropped for convenience) as,
\beq
F_{jk}(q) = F_{jk}^{(s)}(q)F_{jk}^{(a)}(q)
\eeq
where $F_{jk}^{(s)}(q)$ contains the stable roots of $F_{jk}(z)$ and $F_{jk}^{(a)}(q)$ contains the anti-stable roots of $F_{jk}(z)$, which are given by
\beqr
F_{jk}^{(s)}(q) &=& 1 + {f_{jk}}_1^{(s)}q^{-1} + \dots + {f_{jk}}_{n_f}^{(s)}q^{-n_{f_{jk}}^{(s)}}\\
F_{jk}^{(a)}(q) &=& 1 + {f_{jk}}_1^{(a)}q^{-1} + \dots + {f_{jk}}_{n_f}^{(a)}q^{-n_{f_{jk}}^{(a)}}.
\eeqr
We introduce $F_{jk}^{\ast(a)}(q)$ as the monic polynomial whose roots are the mirrored (and stable) roots of $F_{jk}^{(a)}(q)$. We can write $F_{jk}^{\ast(a)}(q)$ as,
\beqr
F_{jk}^{\ast(a)}(q) = 1 + \frac{{f_{jk}}_{n_f-1}^{(a)}}{{f_{jk}}_{n_f}^{(a)}}q^{-1} + \dots + \frac{1}{{f_{jk}}_{n_f}^{(a)}}q^{-n_{f_{jk}}},
\eeqr
assuming without loss of generality that ${f_{jk}}_{n_f}^{(a)} \neq 0$.
Then, we define $F_a(q)$ as the product of all polynomials with anti-stable roots i.e. $F_a(q) = \prod_{k \in \mN_j}F_{jk}^{(a)}(q) = 1 + {f_{a}}_1q^{-1} + \dots + {f_{a}}_{n_f}q^{-n_{f_{a}}}$, and $F_a^\ast(q)$ as the polynomial with mirrored roots of $F_a(q)$ inside the unit circle i.e. $F_a^\ast(q) = \prod_{k \in\mN_j}F_{jk}^{\ast(a)}(q) = 1 + \frac{{f_{a}}_{n_f-1}}{{f_{a}}_{n_f}}q^{-1} + \dots + \frac{1}{{f_{a}}_{n_f}}q^{-n_{f_{a}}}$.

As the next step, we re-write the noise term $v_j(t)$ in \eqref{eq:singleblock} using a the input white noise process $\bar e_j(t)$ instead of $e_j(t)$. Using the fact that $\frac{F_a^\star(q)}{F_a(q)}$ is an all pass filter (linear) with a magnitude of $|\frac{1}{{f_{a}}_{n_f}}|$\cite{Forssell&Ljung_CDC:1998}, we can write $v_j(t) = H_j(q)\frac{F_a^\star(q)}{F_a(q)}\bar e_j(t)$ whose noise spectrum $\Phi_{v_j}$ equals ${|H(e^{i\omega})|}^2  {|\frac{1}{{f_{a}}_{n_f}}|}^2\bar\sigma_j^2$, where $\bar\sigma_j^2={|{f_{a}}_{n_f}|}^2\sigma_j^2$ is the variance of $\bar e_j(t)$.

With the above expression of the noise term and using $G_{ji}(q) = \frac{B_{ji}(q)}{F_{ji}(q)} = \frac{B_{ji}(q)}{F_{ji}^{(s)}(q)F_{ji}^{(a)}(q)}$, and assuming $r_j(t) = 0$ for the sake of brevity, Eq. \eqref{eq:singleblock} is rewritten as,
\begin{align} \label{eq:401}
&w_j(t) = {M_j(q)}w_j(t) - (1-M_j(q)){\bar F_{ji}(q)}w_j(t) \nonumber\\
&+ (1-M_j(q)){B_{ji}(q)}w_i(t) +\!\!\!\!\!\! \sum_{k \in \mathcal{N}_j\backslash\{i\}} \!\!\!\!M_{jk}(q)w_k(t) + \bar e_j(t)
\end{align}
with
\beqr
{M_j(q)} &:=& 1 - \bigg((H_j)^{-1}(q)\frac{\prod_{k\in \mN_j \backslash \{i\}}F_{jk}^{(a)}(q)}{F_a^\star(q)F_{ji}^{(s)}(q)}\bigg) \quad,\\
M_{jk}(q) &:=& (H_j)^{-1}\frac{\prod_{\ell\in \mN_j \backslash \{ k\}}F_{j\ell}^{(a)}(q)}{F_a^\star(q)}\frac{B_{jk}(q)}{F_{jk}^{(s)}(q)} \,,
\eeqr
where $\bar F_{ji}(q) = {-(1 - F_{ji}(q))}$, and $M_j(q)$ is a strictly proper stable filters with only stable poles which are the roots of $F_a^\ast(z)$, $F_{ji}^{(s)}(z)$ and poles of $(H_j)^{-1}$, while $M_{jk}(q), k \in \mN_j\backslash\{i\}$ are also strictly proper stable filters with only stable poles which are the roots of $F_a^\ast(z)$, $F_{jk}^{(s)}(z)$ and poles of $(H_j)^{-1}$.


\section{Proof of Lemma \ref{lemma1}}    
Using the Bayes' rule the expression in Eq.~\eqref{eq:4.1} can be written as,
\begin{equation}\label{eq:5.1}
\begin{split}
&Q^{(n)}(\eta) = \mathbb{E}[{\log p(w_j | m_j, {m_{k}}_1, {m_{k}}_2, \dots, {m_{k}}_p;\eta)}]\\
&+ \mathbb{E}[{\log p(m_j;\eta) + \log p({m_{k}}_1;\eta) + \dots + \log p({m_{k}}_p;\eta)}]
\end{split}
\end{equation}
\begin{equation}
\begin{split}
Q^{(n)}(\eta) = \mathbb{E}[{\mathcal{A}}] + \mathbb{E}[{\mathcal{B}}]
\end{split}
\end{equation}
\begin{equation}
\begin{split}
\mathcal{A} &\coloneqq -\frac{N}{2}\log(2\pi) - \frac{N}{2}\log(\bar\sigma_j^2) -\\
&\frac{1}{2\bar\sigma_j^2}(w_j - \mathbf W_{ji}\bg_{ji} - \mathbf W\bm)^\top(w_j - \mathbf W_{ji}g_{ji} - \mathbf W\bm)
\end{split}
\end{equation}
\begin{equation}\label{eq:9}
\begin{split}
\mathcal{B} \!&\coloneqq \!\!-\frac{l}{2}\log(2\pi) \!\!-\!\! \frac{1}{2}\log[\det({\lambda_j}{{K_\beta}_j})] \!\!-\! \frac{1}{2}{m_{j}}\!^\top\!{({\lambda_j}{{K_\beta}_j})}\!^{-1}{m_{j}} \\
&+ \sum_{k \in \mathcal{N}_j \backslash \{i \}} \bigg[-\frac{l}{2}\log(2\pi) - \frac{1}{2}\log[\det({\lambda_k}{{K_\beta}_k})]\\
&- \frac{1}{2}{m_{k}}^\top{({\lambda_k}{{K_\beta}_k})}^{-1}{m_{k}} \bigg]\\
\end{split}
\end{equation}
\noindent Taking Expectation of each element in $\mathcal{A}$ and $\mathcal{B}$ with respect to $p(\bm|w_j;\hat\eta^{(n)})$ (i.e. $\mathbb{E}_{p(\bm|w_j;\hat\eta^{(n)})}$) we get,
\begin{equation}\label{eq:7.1}
\begin{split}
\mathbb{E}[\mathcal{A}] \!&=\! \!-\frac{N}{2}\!\log(2\pi) \!\!-\! \frac{N}{2}\!\log(\bar\sigma_j^2) \!\!-\! \frac{1}{2\bar\sigma_j^2}\bigg[\!w_{j}^\top w_{j} \!-\! \bg_{ji}^\top \mathbf W_{ji}^\top w_j \\
&- \mathbb{E}[\bm^\top]\mathbf W^\top w_j - w_{j}^\top \mathbf W_{ji}\bg_{ji} + \bg_{ji}^\top \mathbf W_{ji}^\top \mathbf W_{ji}\bg_{ji} +\\ &\mathbb{E}[\bm^\top] \mathbf W^\top \mathbf W_{ji}\bg_{ji} \!-\! w_{j}^\top \mathbf W\mathbb{E}[\bm] + \bg_{ji}^\top \mathbf W_{ji}^\top \mathbf W\mathbb{E}[\bm]\\ 
&+ \mathrm{tr}(\mathbf W^\top \mathbf W\mathbb{E}[\bm\bm^\top]) \bigg]
\end{split}
\end{equation}
\begin{equation}\label{eq:11.1}
\begin{split}
\mathbb{E}[\mathcal{B}] &= -\frac{l}{2}\log(2\pi) - \frac{1}{2}\log[\det({\lambda_j}{{K_\beta}_j})] -\\
&\frac{1}{2}\mathrm{tr}\big({({\lambda_j}{{K_\beta}_j})}^{-1}\mathbb{E}[{m_{j}}{m_{j}}^\top]\big)\\
&+ \sum_{k \in \mathcal{N}_j \backslash \{i \}} \bigg[-\frac{l}{2}\log(2\pi) - \frac{1}{2}\log[\det({\lambda_k}{{K_\beta}_k})]\\
&- \frac{1}{2}\mathrm{tr}\big({({\lambda_k}{{K_\beta}_k})}^{-1}\mathbb{E}[{m_{k}}{m_{k}}^\top]\big)\bigg]\\
\end{split}
\end{equation}
The constants can be removed from the objective functions and multiplication with scalar value 2 can be done to simplify the objective function. On substituting the expected values $\mathbb{E}[\bm\bm^\top] = \hat{\mathbf M}^{(n)}$, $\mathbb{E}[{m_{k}}{m_{k}}^\top] = \hat{\mathbf M}_k^{(n)}$, $\mathbb{E}[{m_{j}}{m_{j}}^\top] = \hat{\mathbf M}_j^{(n)}$ and $\mathbb{E}[\bm] = \hat{\bm}^{(n)}$ we get the statement of the Lemma.

\section{Proof of Theorem \ref{theorem1}}    
The proof follows the procedure used in \cite{Bottegaletal_Autom:16}. We partially differentiate \eqref{eq:102} with respect to $\lambda_{k}$ and equate to zero to get the $\lambda_{k}^*$ expression. Substituting this $\lambda_{k}^*$ in \eqref{eq:102} we get the expression for \eqref{eq:101} using which we obtain ${\hat\beta}_{{k}}^{(n+1)}$. Equation~\eqref{eq:13.1} is the expression of $\lambda_{k}^*$ after substituting ${\hat\beta}_{{k}}^{(n+1)}$.

\section{Proof of Theorem \ref{theorem2}}    
In order to find ${\hat\theta}^{(n)}$, $\bar \sigma_j^2$ is fixed to $\hat{\bar\sigma}_j^{2(n)}$ and substituted in Eq.~\eqref{eq:14.1}. After substitution the terms that are independent of $\theta$ can be removed from the objective function since it becomes a constant. Then we get,
\begin{equation}\label{eq:15.1}
\begin{split}
&Q_o^{(n)}(\theta,\hat{\bar\sigma}_j^{2(n)}) = \mathrm{constant} \quad -\\
&\frac{1}{\hat{\bar\sigma}_j^{2(n)}}\bigg[- 2w_{j}^\top \mathbf W\hat{\bm}^{(n)} + \mathrm{tr}\big(\mathbf W^\top \mathbf W\hat{\mathbf M}^{(n)}\big)\\
&- 2w_{j}^\top \mathbf W_{ji}\bg_{ji} + \bg_{ji}^\top \mathbf W_{ji}^\top \mathbf W_{ji}\bg_{ji} + 2\bg_{ji}^\top \mathbf W_{ji}^\top \mathbf W\hat{\bm}^{(n)}\bigg]\,.
\end{split}
\end{equation}
We know introduce the following notation. Let $D_1 \in \mathbb{R}^{N^2\times N}$ and $D_2 \in \mathbb{R}^{N^2\times N}$ are two matrices such that, for any vector $\mathbf w \in\mathbb{R}^N$, $D_1\mathbf w = \mathrm{vec}(W)$, where $W$ is the Toeplitz matrix of $\mathbf w$, and $D_2\mathbf w = \mathrm{vec}(W^\top)$. Let us define $\breve{m}^{(n)} \in \mathbb{R}^{N}$ be a vector such that, if $N \leq l$, $\breve{m}^{(n)}$ is the vector of first $N$ elements of $\hat m^{(n)}$ and if $N > l$, $\breve{m}^{(n)}$ is a vector with the first $l$ elements equal to $\hat m^{(n)}$ and the remaining ones equal to 0. Let $\breve{M}^{(n)}$, $\overleftrightarrow{W}_\ell^N \in \mathbb{R}^{N\times N}$ where $\ell \in \{i,j\}$ be the Toeplitz matrix of $\breve m^{(n)}$ and $\overleftrightarrow{{w}}_\ell$ respectively.
Then
$$\mathcal{\mathbf X} = {\begin{bmatrix}
	W_j & {W_{k}}_1 & \dots & {W_{k}}_p
	\end{bmatrix}}, \hspace{10pt} \hat{\mathcal{\mathbf Y}}^{(n)} = \breve{M}^{(n)}[\begin{matrix} \overleftrightarrow{W}_i^N  & -\overleftrightarrow{W}_j^N \end{matrix}]$$ and
$$\mathcal{\mathbf Z}_i = {\begin{bmatrix}
	\overleftrightarrow W_{i} & \boldsymbol 0 & \boldsymbol 0 & \dots & \boldsymbol 0
	\end{bmatrix}} \in \mathbb{R}^{N\times (p+1)l} \ , $$
$$\mathcal{\mathbf Z}_j = {\begin{bmatrix}
	-\overleftrightarrow W_{j} & \boldsymbol 0 & \boldsymbol 0 & \dots & \boldsymbol 0
	\end{bmatrix}} \in \mathbb{R}^{N\times (p+1)l} \ . $$
We can re-write the following terms,
$\mathbf W\hat \bm^{(n)} = \mathcal{\mathbf X}\hat \bm^{(n)} +  G_{b}\overleftrightarrow{W}_{i}\hat \bm_j^{(n)} - G_{f}\overleftrightarrow{W}_{j}\hat \bm_j^{(n)}= \mathcal{\mathbf X}\hat \bm^{(n)} + \hat{\mathcal{\mathbf Y}}^{(n)}\bg_{ji}$ and $\mathbf W = \mathcal{\mathbf X} + G_b \mathcal{\mathbf Z}_i + G_f \mathcal{\mathbf Z}_j$.
Therefore,
\begin{equation*}
\begin{split}
&\hat{\theta}^{(n+1)}\\
&= \argmax_\theta \bigg[2w_{j}^\top \mathbf W\hat{\bm}^{(n)} - \mathrm{tr}\big(\mathbf W^\top \mathbf W\hat{\mathbf M}^{(n)}\big)\\ 
&+ 2w_{j}^\top \mathbf W_{ji}\bg_{ji} - \bg_{ji}^\top \mathbf W_{ji}^\top \mathbf W_{ji}\bg_{ji} - 2\bg_{ji}^\top \mathbf W_{ji}^\top \mathbf W\hat{\bm}^{(n)}\bigg]\\
&= \argmax_\theta \bigg[2w_{j}^\top\mathbf X\hat{\bm}^{(n)} \!+\! 2w_{j}^\top\hat{\mathbf  Y}^{(n)}{\bg}_{ji} \!-\! \mathrm{tr}\big(\mathbf{XX^\top}\hat{\mathbf M}^{(n)}\big)\\
&-\mathrm{tr}\big(\mathbf X\hat{\mathbf M}^{(n)}\mathbf{Z}_j^\top G_{f}^\top\big) - \mathrm{tr}\big(\mathbf Z_i\hat{\mathbf M}^{(n)}\mathbf{X^\top}G_{b}\big) - \\ &\mathrm{tr}\big(\mathbf Z_j\hat{\mathbf M}^{(n)}\mathbf{X^\top}G_{f}\big) - \mathrm{tr}\big(G_{b}\mathbf Z_i\hat{\mathbf M}^{(n)}\mathbf{Z}_i^\top G_{b}^\top \big) - \\ &\mathrm{tr}\big(G_{f}\mathbf Z_j\hat{\mathbf M}^{(n)}\mathbf{Z}_j^\top G_{f}^\top \big) - \mathrm{tr}\big(G_{b}\mathbf Z_i\hat{\mathbf M}^{(n)}\mathbf {Z}_j^\top G_{f}^\top \big) - \\ &\mathrm{tr}\big(G_{f}\mathbf Z_j\hat{\mathbf M}^{(n)}\mathbf {Z}_i^\top G_{b}^\top \big) - \mathrm{tr}\big(\mathbf X\hat{\mathbf M}^{(n)}\mathbf {Z}_i^\top G_{b}^\top\big)\\
&+ 2w_{j}^\top \mathbf W_{ji}\bg_{ji} - \bg_{ji}^\top \mathbf W_{ji}^\top \mathbf W_{ji}\bg_{ji} \\
&- 2\bg_{ji}^\top \mathbf W_{ji}^\top \mathbf X\hat{\bm}^{(n)} - 2\bg_{ji}^\top \mathbf W_{ji}^\top \hat{\mathbf Y}^{(n)}\bg_{ji}\bigg]\\
\end{split}
\end{equation*}
Neglecting constant terms we get,
\begin{equation*}
\begin{split}
&\hat{\theta}^{(n+1)}=\\
&= \argmax_\theta \bigg[2w_{j}^\top\hat{\mathbf Y}^{(n)}{\bg}_{ji} - \mathrm{tr}\big(\mathbf X\hat{\mathbf M}^{(n)}\mathbf{Z}_i^\top G_{b}^\top\big) - \\
&\mathrm{tr}\big(\mathbf X\hat{\mathbf M}^{(n)}\mathbf {Z}_j^\top G_{f}^\top\big) - \mathrm{tr}\big(\mathbf Z_i\hat{\mathbf M}^{(n)}\mathbf {X^\top}G_{b}\big) - \\ &\mathrm{tr}\big(\mathbf Z_j\hat{\mathbf M}^{(n)}\mathbf {X^\top}G_{f}\big) - \mathrm{tr}\big(G_{b}\mathbf Z_i\hat{\mathbf M}^{(n)}\mathbf {Z}_i^\top G_{b}^\top \big) - \\ &\mathrm{tr}\big(G_{f}\mathbf Z_j\hat{\mathbf M}^{(n)}\mathbf {Z}_j^\top G_{f}^\top \big) - \mathrm{tr}\big(G_{b}\mathbf Z_i\hat{\mathbf M}^{(n)}\mathbf {Z}_j^\top G_{f}^\top \big) - \\ &\mathrm{tr}\big(G_{f}\mathbf Z_j\hat{\mathbf M}^{(n)}\mathbf {Z}_i^\top G_{b}^\top \big) + 2w_{j}^\top \mathbf W_{ji}\bg_{ji}-\\
&\bg_{ji}^\top \mathbf W_{ji}^\top \mathbf W_{ji}\bg_{ji} \!-\! 2\hat{\bm}^{(n)\top}\mathbf X^\top \mathbf W_{ji}\bg_{ji} \!-\! 2\bg_{ji}^\top \mathbf W_{ji}^\top \hat{\mathbf Y}^{(n)}\bg_{ji}\bigg]
\end{split}
\end{equation*}
\begin{equation*}
\begin{split}
&= \argmax_\theta \bigg[2w_{j}^\top\hat{\mathbf Y}^{(n)}{\bg}_{ji} - \mathrm{vec}(\mathbf {Z}_i\hat{\mathbf M}^{(n)\top}\mathbf X^\top)^\top D_2b_{ji} -\\
&\mathrm{vec}(\mathbf {Z}_j\hat{\mathbf M}^{(n)\top}\mathbf X^\top)^\top D_2f_{ji}
- \mathrm{vec}(\mathbf X\hat{\mathbf M}^{(n)\top}\mathbf {Z}_i^\top)^\top D_1b_{ji} \\
&- \mathrm{vec}(\mathbf X\hat{\mathbf M}^{(n)\top}\mathbf {Z}_j^\top)^\top D_1f_{ji} + 2w_{j}^\top \mathbf W_{ji}\bg_{ji} \\
&- b_{ji}^\top D_1^\top(\mathbf {Z}_i\hat{\mathbf M}^{(n)}\mathbf {Z}_i^\top \otimes I_N)D_1b_{ji}\\
&- f_{ji}^\top D_1^\top(\mathbf {Z}_j\hat{\mathbf M}^{(n)}\mathbf {Z}_j^\top \otimes I_N)D_1f_{ji}\\
&- b_{ji}^\top D_1^\top(\mathbf {Z}_i\hat{\mathbf M}^{(n)}\mathbf {Z}_j^\top \otimes I_N)D_1f_{ji}\\
&- f_{ji}^\top D_1^\top(\mathbf {Z}_j\hat{\mathbf M}^{(n)}\mathbf {Z}_i^\top \otimes I_N)D_1b_{ji}\\
&- \bg_{ji}^\top \mathbf W_{ji}^\top \mathbf W_{ji}\bg_{ji} - 2\hat{\bm}^{(n)\top}\mathbf X^\top \mathbf W_{ji}\bg_{ji}\\
&- 2\bg_{ji}^\top \mathbf W_{ji}^\top \hat{\mathbf Y}^{(n)}\bg_{ji}\bigg].
\end{split}
\end{equation*}
Defining
\beqr
\hat A_{11}^{(n)} & = & [D_1^\top(\mathbf {Z}_i\hat{\mathbf M}^{(n)}\mathbf {Z}_i^\top \otimes I_N)D_1] \nonumber\\
\hat A_{12}^{(n)} & = & [D_1^\top(\mathbf {Z}_i\hat{\mathbf M}^{(n)}\mathbf {Z}_j^\top \otimes I_N)D_1] \nonumber\\
\hat A_{21}^{(n)} & = & [ D_1^\top(\mathbf {Z}_j\hat{\mathbf M}^{(n)}\mathbf {Z}_i^\top \otimes I_N)D_1] \nonumber\\
\hat A_{22}^{(n)} & = & [D_1^\top(\mathbf {Z}_j\hat{\mathbf M}^{(n)}\mathbf {Z}_j^\top \otimes I_N)D_1] \nonumber
\eeqr
\beq
\begin{split}
	\hat{b}_{11}^{(n)} & =  \Big[- \frac{1}{2}\mathrm{vec}(\mathbf {Z}_i\hat{\mathbf M}^{(n)\top}\mathbf X^\top)^\top D_2\\ & \quad \quad \quad \quad - \frac{1}{2}\mathrm{vec}(\mathbf X\hat{\mathbf M}^{(n)\top}\mathbf {Z}_i^\top)^\top D_1\Big]^\top, \nonumber\\
	\hat{b}_{12}^{(n)} & =  \Big[- \frac{1}{2}\mathrm{vec}(\mathbf {Z}_j\hat{\mathbf M}^{(n)\top}\mathbf X^\top)^\top D_2\\ & \quad \quad \quad \quad - \frac{1}{2}\mathrm{vec}(\mathbf X\hat{\mathbf M}^{(n)\top}\mathbf {Z}_j^\top)^\top D_1\Big]^\top \nonumber
\end{split}
\eeq
and
\beq
\begin{split}
	\hat{\mathbf A}^{(n)} & = \begin{bmatrix}
		\hat A_{11}^{(n)} & \hat A_{12}^{(n)} \\
		\hat A_{21}^{(n)} & \hat A_{22}^{(n)}
	\end{bmatrix} + \mathbf W_{ji}^\top \mathbf W_{ji} + 2\mathbf W_{ji}^\top \hat{\mathbf Y}^{(n)} \ , \nonumber \\
	\hat{\mathbf b}^{(n)} & = \begin{bmatrix}
		\hat b_{11}^{(n)} \\ \hat b_{12}^{(n)}
	\end{bmatrix} + [\begin{matrix} w_{j}^\top \mathbf W_{ji} + w_{j}^\top\hat{\mathbf Y}^{(n)} - \hat{\bm}^{(n)\top}\mathbf X^\top \mathbf W_{ji}
	\end{matrix}]^\top \nonumber
\end{split}
\eeq
we get that the parameter vector $\theta$ are updated by solving the problem
\begin{equation}
\hat{\theta}^{(n+1)} = \argmin_\theta \bigg[\bg_{ji}^\top \hat{\mathbf A}^{(n)}\bg_{ji} - 2\hat{\mathbf b}^{(n)\top}\bg_{ji}\bigg].
\end{equation}
We have $\bg_{ji}$ to be linearly parameterized with $\theta$, that is $\bg_{ji} = M\theta$ where $M \in \mathbb{R}^{2N\times n_{\theta}}$. Therefore, the above problem becomes quadratic and a closed-form solution is achieved. Thus we get the statement of Theorem \ref{theorem2}.

\section{Proof of Theorem \ref{theorem3}}    
In order to find $\hat{\bar\sigma}_j^{2(n)}$, $\theta$ is fixed to ${\hat\theta}^{(n+1)}$ and substituted in Eq.~\eqref{eq:14.1}. After substitution, $Q_o^{(n)}(\bar\sigma_j^2,\hat{\theta}^{(n+1)})$ is differentiated w.r.t. $\bar\sigma_j^2$ and equated to zero to get the statement of the Theorem.

\end{document}